\documentclass[trackchanges, twocolumn, twocolappendix]{aastex701}
\usepackage{amsmath}

\newcommand{\mytilde}{\raise.19ex\hbox{$\scriptstyle\sim$}}
\newcommand{\sersic}{S\'ersic }

\newcommand{\sma}{\mathrm{SMA}}
\shorttitle{MUSE BCG-ICL Decomposition: AS1063, MACS1206}
\shortauthors{Joo and Jee}

\received{}
\revised{}
\accepted{}%

\submitjournal{ApJ}

\begin{document}

\title{Kinematic Decomposition of the BCG and ICL in Abell S1063 and MACS J1206}

\correspondingauthor{M. James Jee}
\email{gudwls4478@yonsei.ac.kr, mkjee@yonsei.ac.kr}

\author[orcid=0000-0001-9139-5455,gname=Hyungjin, sname=Joo]{Hyungjin Joo} 
\affiliation{
Department of Astronomy,
Yonsei University, 50 Yonsei-ro, Seoul 03722, Republic of Korea}
\email{gudwls4478@gmail.com}

\author[orcid=0000-0002-5751-3697,gname=Myungkook James, sname=Jee]{M. James Jee} 
\affiliation{
Department of Astronomy,
Yonsei University, 50 Yonsei-ro, Seoul 03722, Republic of Korea}
\affiliation{
Department of Physics,
University of California, Davis, One Shields Avenue, Davis, CA 95616, USA}
\email{mkjee@yonsei.ac.kr}

\begin{abstract}
The brightest cluster galaxy (BCG) and the intracluster light (ICL) overlap in space and merge smoothly in surface brightness.
Beyond the nearest clusters, measurements of the ICL have therefore relied on photometric definitions.
We present a spectral decomposition of the light in cluster cores into BCG and ICL, using deep VLT/MUSE integral-field spectroscopy of Abell~S1063 ($z=0.35$) and MACS~J1206.2$-$0847 ($z=0.44$).
Spectral templates of the two components are derived iteratively from the BCG center and the ICL-dominated outskirts, and their weighted sum is fitted to the spectrum of every spatial bin.
This enables us to measure the kinematics and the light of the BCG and ICL separately without assuming a boundary or a functional form for their light profiles.
The velocity dispersions of the BCG and ICL are $316\pm5$ and $753\pm25~\mathrm{km~s^{-1}}$, respectively, in AS1063, whereas we obtain $313\pm5$ and $695^{+47}_{-45}~\mathrm{km~s^{-1}}$ in MACS1206. The ICL is blueshifted relative to the BCG by $75\pm22$ and $148\pm38~\mathrm{km~s^{-1}}$, respectively, and is $\mytilde0.7$~dex more metal-poor, at a combined statistical significance of $\mytilde2\sigma$.
Defining the ICL as the light fainter than $\mu_{B}=25~\mathrm{mag~arcsec^{-2}}$ assigns two-thirds to three-quarters of the kinematically defined ICL to the BCG, while a double \sersic model fits the total profiles well but leaves the division between the two components poorly constrained given the depth and resolution of the current MUSE data.
Our kinematic decomposition provides a direct, physically motivated reference for photometric definitions of the ICL.
\end{abstract}


\section{Introduction} \label{sec:intro}
Since \citet{Zwicky1951} first reported faint diffuse light between galaxies in the Coma cluster, numerous studies have explored its nature and origin.
This faint light, originating from stars that are not gravitationally bound to individual galaxies, has been referred to as intracluster light (ICL).
The ICL is usually observed as an extended, low-surface-brightness feature around the brightest cluster galaxy (BCG).

The ICL has the potential to trace the dark matter (DM) in clusters, as suggested by both observational \citep[e.g.,][]{Jee2010, Montes2019, Cha2025} and theoretical \citep[e.g.,][]{AlonsoAsensio2020, Yoo2024} studies.
Because it accumulates over a wide range of epochs, the ICL also maintains a record of the assembly history of the cluster \citep[e.g.,][]{Contini2021, Montes2022, Joo2026}.
However, these applications require that the ICL be measured accurately, both in its amount and in its spatial distribution.

Accurate measurement of the ICL is challenging because the ICL and the BCG are physically intertwined.
The ICL can be produced by the tidal stripping of satellite galaxies \citep[e.g.,][]{Rudick2009, Contini2014} and by pre-processing in galaxy groups prior to their accretion onto the cluster \citep[e.g.,][]{Willman2004, Rudick2006}.
In addition, the ICL can be produced during the formation of the BCG, as stars become unbound through mergers or violent relaxation in the central region of the cluster \citep[e.g.,][]{Murante2007, Conroy2007, Ko2018, Joo2025}.
Tidal stripping preferentially removes stars from low-mass satellites and from the outskirts of massive ones, both of which are relatively metal-poor.
This predicts an ICL that is more metal-poor than the BCG, and the negative color and metallicity gradients observed in BCG+ICL systems support this expectation \citep[e.g.,][]{Montes2014, DeMaio2015, Morishita2017, DeMaio2018, Montes2018}.
The stripped stars settle near the outskirts of the BCG, where the BCG and the inner ICL overlap.
As a result, the surface brightness profile varies smoothly from the BCG to the ICL.
The light profile alone provides no physical criterion for where the BCG ends and the ICL begins.

Observational studies have defined the ICL as the light fainter than a surface brightness threshold \citep[e.g.,][]{Feldmeier2004, Burke2015, Montes2018} or as the light beyond a certain physical radius from the BCG \citep[e.g.,][]{Pillepich2018}.
Other studies have modeled the total light profile with multiple components and assigned the outer component to the ICL \citep[e.g.,][]{Gonzalez2005, Zibetti2005, Kluge2020, Joo2023, Joo2026}.
These definitions use different assumptions, and the ICL fraction of the same cluster can vary by a factor of several depending on the definition adopted \citep[e.g.,][]{Rudick2011, Kluge2021, Montes2022}.
Moreover, semi-analytic models suggest that the boundary between the two components is neither a fixed surface brightness nor a fixed radius, but varies from system to system \citep{Contini2022, Contini2026}. If so, a single definition applied to different clusters would misclassify different fractions of the ICL.
Testing any definition requires a reference that separates the two components without depending on the light profile.

The stellar kinematics provide such a criterion.
In numerical simulations, the stars of the BCG and those of the ICL show distinct velocity distributions, and these differences have been used to separate the two components \citep[e.g.,][]{Dolag2010, Cui2014, Remus2017, Marini2024}.
In nearby clusters, planetary nebulae have been used as discrete tracers of the diffuse light, revealing that the stellar halo of the BCG and the ICL have distinct kinematics \citep[e.g.,][]{Longobardi2015, Hartke2018}.
However, this technique is limited to distances of $\lesssim100$~Mpc.
For more distant clusters, long-slit and integral-field spectroscopy (IFS) have revealed velocity dispersion profiles that increase from the BCG toward the outskirts \citep[e.g.,][]{Dressler1979, Kelson2002, Loubser2008, Newman2013, Loubser2018, Edwards2020}.
In a few nearby cD galaxies, the rising dispersion has been interpreted as the signature of a halo bound to the cluster rather than to the galaxy \citep{Bender2015}, and multiple kinematic components have been identified in the halo of NGC~3311 \citep{Barbosa2018}, whose rising dispersion has been interpreted as a superposition of tracer populations with different spatial distributions \citep{Hilker2018}.
Beyond the nearest clusters, however, the stellar light has been described with a single velocity dispersion.
These measurements provide neither the kinematics of the two components separately nor the light distribution of each component.
This mixture also affects studies that use the kinematics of the BCG to constrain the inner mass distribution of clusters, where the BCG stars are treated as a single tracer \citep[e.g.,][]{Newman2013}.

In this study, we separate the BCG and ICL by their stellar kinematics based on deep IFS observations of two massive galaxy clusters, AS1063 and MACS1206, obtained with the Multi Unit Spectroscopic Explorer (MUSE) on the Very Large Telescope (VLT).
This approach requires neither a boundary between the two components nor a functional form for their light profiles, and thus provides a physically motivated reference against which the results from the conventional definitions can be compared.

While this paper was being finalized, \citet{Bellhouse2026} presented a kinematic decomposition of the BCG and ICL in AS1063 with MUSE, whose preliminary results\footnote{Tracing Cosmic Evolution with Galaxy Clusters V, Sexten Center for Astrophysics, Sesto, Italy, 2025 July 7--11; \url{https://www.sexten-cfa.eu/event/sestocluster2025/}.} motivated this work.
Our analysis was carried out independently, using MUSE data of AS1063 with about four times their integration time and including an additional cluster, MACS1206.

Our paper is organized as follows.
In \textsection~\ref{sec:data}, we describe the MUSE observations and the data reduction.
\textsection~\ref{sec:method} presents our method for decomposing the spectra into the BCG and ICL using their optimal stellar templates.
In \textsection~\ref{sec:result}, we present the kinematic and surface brightness profiles of the two components.
In \textsection~\ref{sec:discussion}, we use the kinematic decomposition to investigate the conventional definitions of the BCG--ICL boundary and the multi-component profile fitting, discuss the origin of the rising velocity dispersion profiles of BCGs, and compare our results with those of \citet{Bellhouse2026}. We conclude in \textsection~\ref{sec:conclusion}.
We report magnitudes in the AB system and adopt a flat $\Lambda$CDM cosmology with $H_{0}=67.7~\mathrm{km\ s^{-1}\ Mpc^{-1}}$ and $\Omega_{m}=0.31$ \citep{Planck2020}.
At the redshifts of AS1063 ($z=0.35$) and MACS1206 ($z=0.44$), this cosmology gives angular scales of $1^{\prime\prime}=5.07$ and $5.86$~kpc, respectively.

\section{Data} \label{sec:data}

\begin{figure*}
    \centering
    \includegraphics[width=0.95\textwidth]{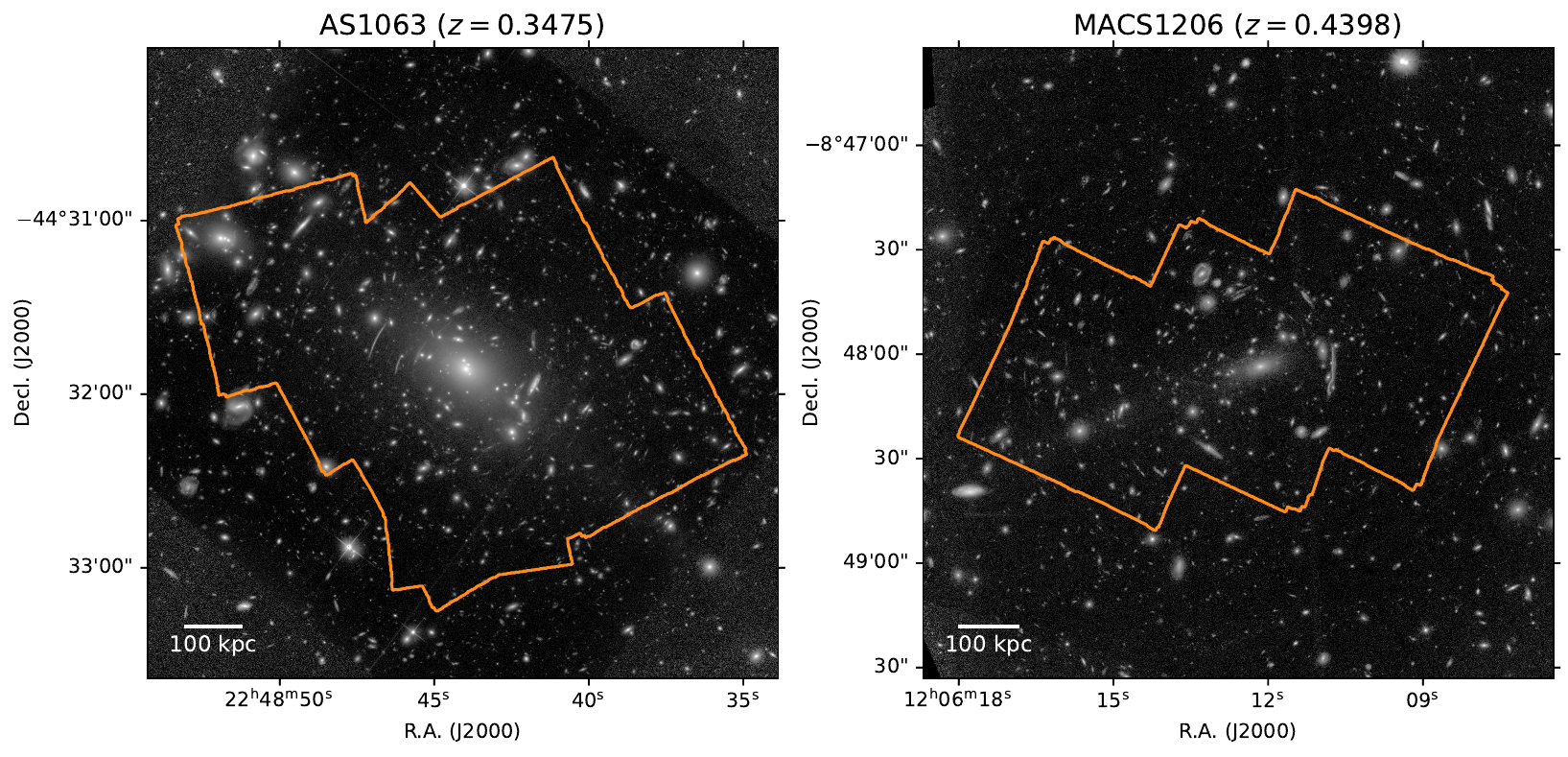}\\[4pt]
    \includegraphics[width=0.95\textwidth]{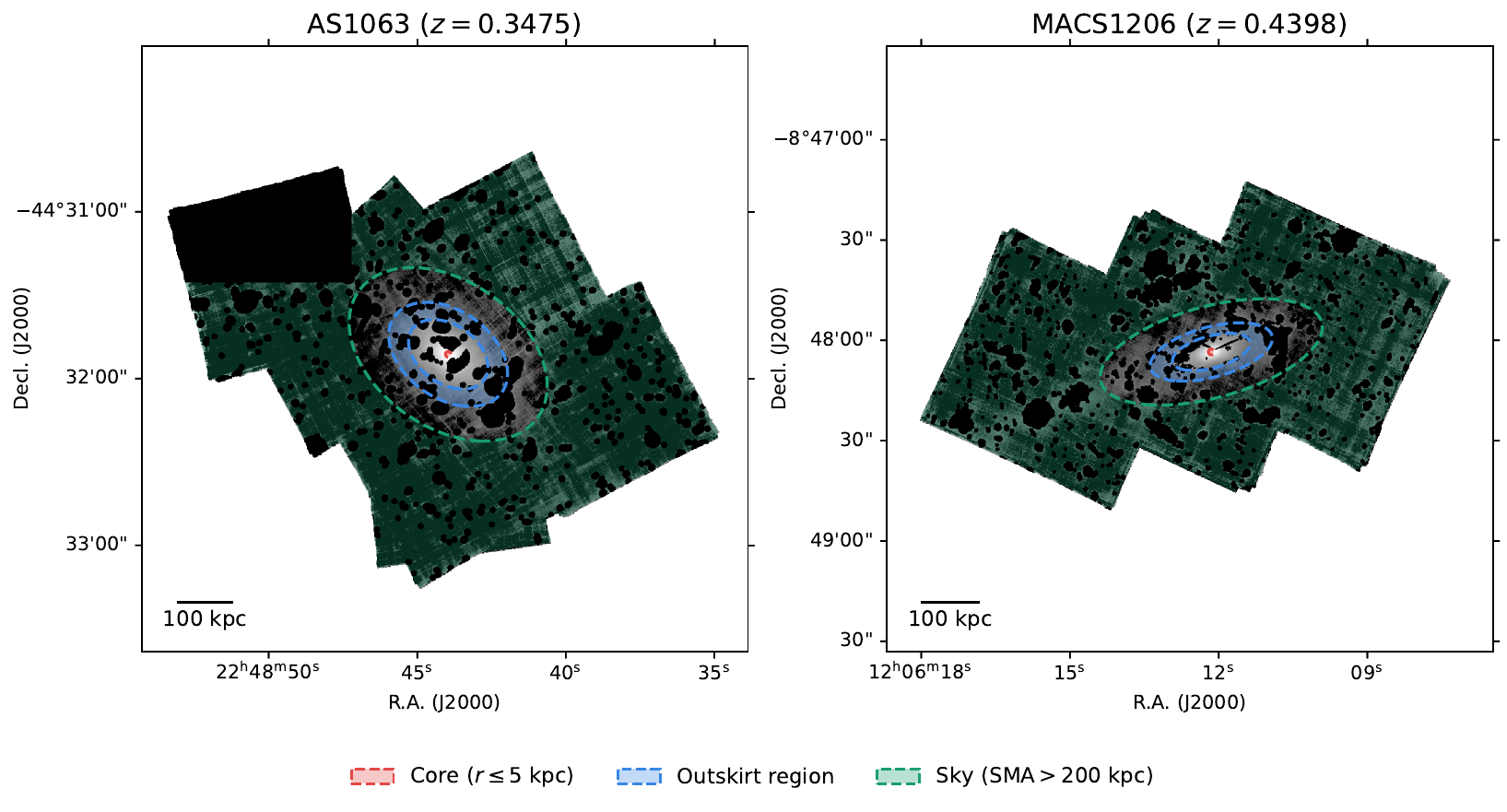}
    \caption{
    Top: VLT/MUSE footprints of AS1063 (left) and MACS1206 (right) overlaid on the HST F606W+F814W images.
    The orange outline marks the region covered by the combined MUSE data. The white bar corresponds to 100~kpc.
    Both footprints extend to $\gtrsim400$~kpc from the BCG along the major axis.
    Bottom: VLT/MUSE images of AS1063 (left) and MACS1206 (right) with the source mask and the regions used in this study.
    Each image is obtained by collapsing the data cube over the rest-frame $3500$--$5500~\text{\AA}$ range, and the black regions are masked based on the HST source detection (\textsection~\ref{subsec:masking}).
    The shaded regions indicate the regions used to construct the BCG and ICL templates (\textsection~\ref{subsec:templates}): the core ($r\leq5$~kpc; red) and the outskirt region ($80\leq\sma\leq120$~kpc for AS1063 and $70\leq\sma\leq110$~kpc for MACS1206; blue).
    The green shaded region ($\sma>200$~kpc) represents the sky area that we used to obtain the sky spectrum.
    }
    \label{fig:footprint}
    \label{fig:masked}
\end{figure*}

\subsection{Targets} \label{subsec:targets}
In this work, we analyze two massive galaxy clusters, Abell~S1063 (hereafter AS1063) and MACS~J1206.2$-$0847 (hereafter MACS1206), observed with MUSE on the VLT.
We selected these targets from the MUSE archive based on two criteria.
First, the redshift should be high enough to shift the Balmer break into the MUSE wavelength range, yet low enough to keep the rest-frame optical continuum of the ICL blueward of the strong sky lines ($\lambda_{\mathrm{obs}} \gtrsim 8000~\text{\AA}$).
This wavelength range brackets the Balmer break and the strongest absorption features in the optical, including Ca~II H and K, the G band, H$\beta$, and the Mg~$b$ triplet. These features anchor the velocity dispersion measurements and constrain the ages and metallicities of the BCG and ICL templates (\textsection~\ref{subsec:templates}).
With the Balmer break at $3646~\text{\AA}$ and the blue limit of MUSE at $4750~\text{\AA}$, the former condition requires $z \gtrsim 0.30$, and keeping the rest-frame continuum up to $5500~\text{\AA}$ below $8000~\text{\AA}$ requires $z \lesssim 0.45$.
Second, among the clusters satisfying this redshift condition, we required multi-pointing mosaics that cover the BCG and its surroundings out to several hundred kpc, with total integration times long enough to detect the ICL.

AS1063 at $z=0.3475$ \citep{Bohringer2004} and MACS1206 at $z=0.4398$ \citep{Biviano2023} satisfy both criteria.
At these redshifts, the Balmer break is observed at $4913$ and $5250~\text{\AA}$, and the rest-frame $3500$--$5500~\text{\AA}$ window used in our fits (\textsection~\ref{subsec:templates}) corresponds to observed wavelengths of $\mytilde4750$--$7410~\text{\AA}$ for AS1063 and $5040$--$7920~\text{\AA}$ for MACS1206.

The two clusters are among the most massive known and were targets of the CLASH program \citep{Postman2012}, and AS1063 was also a target of the Hubble Frontier Fields \citep{Lotz2017}. These programs provide the deep HST imaging used for source masking in this work (\textsection~\ref{subsec:masking}).
AS1063 is the second most X-ray luminous cluster in the REFLEX survey \citep{Bohringer2004}, with a weak-lensing mass of $M_{200m}=3.3^{+1.0}_{-0.7}\times10^{15}~M_{\odot}$ \citep{Gruen2013} and a line-of-sight velocity dispersion of $1380^{+26}_{-32}~\mathrm{km~s^{-1}}$ from 1234 spectroscopic members \citep{Mercurio2021}.
Despite its regular appearance, its velocity distribution is bimodal and its member galaxies are elongated along the NE--SW direction, indicating that the cluster is far from dynamically relaxed \citep{Mercurio2021}.
MACS1206 has a lensing virial mass of $M_{\mathrm{vir}}=(1.1\pm0.2)\times10^{15}~h^{-1}M_{\odot}$ \citep{Umetsu2012}, a dynamical mass of $M_{200}=(1.4\pm0.2)\times10^{15}~M_{\odot}$, and a velocity dispersion of $1087^{+53}_{-55}~\mathrm{km~s^{-1}}$ from $\mytilde600$ members \citep{Biviano2013}.
It appears relaxed in projection, with the X-ray peak coincident with the BCG \citep{Biviano2013}, but its galaxy distribution is elongated along the WNW--ESE direction, interpreted as the main direction of accretion \citep{Girardi2015}, and its ICL shows an asymmetric excess toward the SE \citep{Presotto2014}.
The global ellipses adopted in \textsection~\ref{subsec:sky}, with position angles of $55^{\circ}$ and $105^{\circ}$ for AS1063 and MACS1206, follow these elongations.

\subsection{MUSE Observations} \label{subsec:obs}
All observations were taken in the MUSE wide-field mode (WFM), which provides a $1^{\prime}\times1^{\prime}$ field of view sampled at $0.2^{\prime\prime}~\mathrm{spaxel^{-1}}$ and a spectral sampling of $1.25~\text{\AA}$ over the nominal wavelength range of $\mytilde 4750$--$9350~\text{\AA}$.
The WFM can be operated either without adaptive optics (standard mode) or with the ground-layer adaptive optics system (AO mode).
All data used in this study were taken in the nominal wavelength setting of each mode (WFM-NOAO-N and WFM-AO-N).

The AS1063 data consist of 83 exposures in seven pointings that form an irregular mosaic around the BCG (Figure~\ref{fig:footprint}): a central pointing on the BCG is flanked by pointings along the NE--SW major axis and toward the NW and SE, covering $\mytilde3.4^{\prime}\times2.5^{\prime}$ in total and reaching $\mytilde500$~kpc from the BCG along the major axis and $\mytilde400$~kpc along the minor axis.
Five of these pointings were observed in the AO mode under programs 114.28L1.001 (PI: S. Fujimoto) and 116.28VK.001 (PI: J. Richard), comprising 63 exposures of 1304~s each.
The remaining two pointings were observed in the standard mode under the MUSE Science Verification program 60.A-9345(A) and program 095.A-0653(A) (PI: K. Caputi), comprising 8 exposures of 1420~s and 12 exposures of 1447~s, respectively.
Although the two modes deliver different spatial resolutions, we combined all exposures to maximize the sensitivity to the faint diffuse light, after placing them on a common flux scale (\textsection~\ref{subsec:reduction}).
The total integration time is $\mytilde 30.8$~hr.

The MACS1206 data consist of 26 exposures in three overlapping pointings aligned along the WNW--ESE major axis of the cluster, with the BCG in the central pointing (Figure~\ref{fig:footprint}), covering $\mytilde3^{\prime}\times1.5^{\prime}$ in total and reaching $\mytilde500$~kpc from the BCG along the major axis; all were obtained in the standard mode under programs 095.A-0181(A) and 097.A-0269(A) (PI: J. Richard).
All exposures but one have an exposure time of 1800~s, yielding a total integration time of $\mytilde 12.9$~hr.
The combined data cubes reach $1\sigma$ surface brightness limits of $26.9$ (AS1063) and $27.2~\mathrm{mag~arcsec^{-2}}$ (MACS1206) in the rest-frame $3500$--$5500~\text{\AA}$ band, measured from the spaxel-to-spaxel scatter in the sky region (\textsection~\ref{subsec:sky}).

\subsection{Data Reduction} \label{subsec:reduction}
The raw data were processed with the MUSE pipeline version 2.10.16 \citep{Weilbacher2020} through the ESO Recipe Execution Tool (\texttt{EsoRex}), including bias subtraction, flat-fielding, wavelength calibration, and illumination correction with twilight sky flats.
Because the ICL may extend over a large fraction of the MUSE field of view, even the faintest spaxels in each exposure can contain a non-negligible contribution from the diffuse light.
To avoid any oversubtraction of the ICL, we turned off the sky subtraction of the pipeline and instead estimated the sky from the combined data cube (\textsection~\ref{subsec:sky}).

To correct the astrometric offsets between exposures, we matched the sources in the white-light image of each exposure to those in the HST F606W+F814W mosaic and applied the resulting translational offsets.
Each exposure was then resampled onto a common spatial and spectral grid for each cluster.

The exposures differ in exposure time, observing conditions, and instrument mode, and thus in their background levels.
We converted all exposures into count rates (counts~s$^{-1}$) to account for their different exposure times.
We then matched the background levels of the exposures in two steps, after masking out all sources detected in the HST images with \texttt{SExtractor} \citep{Bertin1996}.
First, within each pointing, we measured the median difference between each exposure and the first exposure of the pointing at each wavelength, using their overlapping spaxels, and subtracted it as an additive offset.
Second, we matched the pointings in the same manner, relative to the central pointing of each cluster.
The matched exposures were combined into a single data cube for each cluster using a $3\sigma$-clipped median.
The variance of each spaxel was propagated through all of the above steps and is used, together with the empirical scatter and the sky uncertainty, to estimate the flux uncertainty (\textsection~\ref{subsec:binning}).

Finally, we converted the count rates into flux density units ($\mathrm{erg~s^{-1}~cm^{-2}~\text{\AA}^{-1}}$) using the master response curves provided with the MUSE pipeline \citep{Weilbacher2020}.
We applied the response curve of the central pointing's mode (AO for AS1063, standard for MACS1206). The two curves agree to within $2\%$ over our wavelength range except at $5800$--$5975~\text{\AA}$ (observed frame), where the AO notch filter blocks the light of the sodium laser guide stars at $\mytilde 5890~\text{\AA}$; there we replaced the AO curve with a linear interpolation, so the choice of curve has a negligible effect on the combined AS1063 data.

\subsection{Source Masking} \label{subsec:masking}
To isolate the diffuse light, we masked out the light from discrete sources other than the BCG.
Because the spatial resolution of the MUSE data is limited by seeing, we constructed the mask from the HST F606W+F814W mosaic, where compact sources are much better resolved.
We started from the \texttt{SExtractor} segmentation map used for the background matching (\textsection~\ref{subsec:reduction}).
To include the faint wings of the sources, we expanded each segment using a binary dilation with a width proportional to the size of the segment, following \citet{Joo2023}, and chose the proportionality factor by visually inspecting the masked collapsed images of the combined data cubes.

We then removed the BCG from the mask to preserve its light, and reprojected the mask onto the MUSE pixel grid.
Finally, we manually masked the sources near the BCG that were not deblended in the segmentation map, and, for AS1063, the extended wings of a bright galaxy at the northeastern edge of its field of view.
The final masked images are shown in Figure~\ref{fig:masked}.
All subsequent measurements, including the $S/N$ maps, spatial binning, and elliptical annuli, are based on the masked data cubes.

\subsection{Elliptical Geometry and Sky Estimation} \label{subsec:sky}
All radial measurements in this work use a single elliptical geometry for each cluster, characterized by the BCG center, position angle (PA), and axis ratio ($Q=b/a$).
We determined the BCG center with the quadratic centroid algorithm in \texttt{photutils} \citep{Bradley2024} on the masked collapsed image.
We then performed isophotal fitting with the \texttt{Ellipse} task in \texttt{photutils} and adopted a global ellipse that represents the overall shape of the BCG+ICL.
The resulting global ellipses have $\mathrm{PA}=55^{\circ}$ and $Q=0.67$ for AS1063, and $\mathrm{PA}=105^{\circ}$ and $Q=0.40$ for MACS1206, where the PA is measured from north to east.
For a spaxel at offset $(\Delta x, \Delta y)$ from the BCG center, we define the elliptical radius as
\begin{equation}
    \sma = \sqrt{x_{\mathrm{rot}}^{2} + \left(\frac{y_{\mathrm{rot}}}{Q}\right)^{2}},
    \label{eq:sma}
\end{equation}
where $x_{\mathrm{rot}}$ and $y_{\mathrm{rot}}$ are the coordinates rotated to align with the major and minor axes of the ellipse, respectively.

We defined the sky region as the spaxels beyond $\sma=200~\mathrm{kpc}$, after excluding all spaxels covered by the source mask (\textsection~\ref{subsec:masking}).
This region contains $\mytilde 7.2\times10^{5}$ (AS1063) and $\mytilde 3.1\times10^{5}$ (MACS1206) spaxels (see Figure~\ref{fig:masked}).
At each wavelength, we adopted the $3\sigma$-clipped median of these spaxels as the sky level and subtracted it from every spaxel.
To account for the systematic uncertainty in the sky estimation, we divided the sky region into 20 azimuthal bins and measured the median sky level within each bin at each wavelength.
We adopted the standard deviation of these 20 values as the sky uncertainty, $\sigma_{\mathrm{sky}}$, and incorporated it into the error budget of all subsequent measurements.

The ICL may extend beyond $\sma=200~\mathrm{kpc}$ and contribute to this sky estimate. However, the median residual in elliptical annuli beyond 200~kpc stays within $0.2\%$ (AS1063) and $0.07\%$ (MACS1206) of the sky level in the rest-frame $3500$--$5500~\text{\AA}$ band and shows no radial trend, so any such contribution is negligible within the precision of our data.

\section{Method} \label{sec:method}
\subsection{BCG and ICL Templates} \label{subsec:templates}
The BCG and ICL stars occupy different orbits in the cluster potential and are thus expected to have distinct line-of-sight velocity distributions (LOSVDs).
We therefore modeled the observed spectrum of each bin as a superposition of two stellar components, the BCG and the ICL, each with its own line-of-sight velocity ($V$) and velocity dispersion ($\sigma$).
We measured the velocities relative to the cluster redshifts, $z=0.3475$ (AS1063) and $0.4398$ (MACS1206; \textsection~\ref{subsec:targets}).

\begin{figure}
    \centering
    \includegraphics[width=\columnwidth]{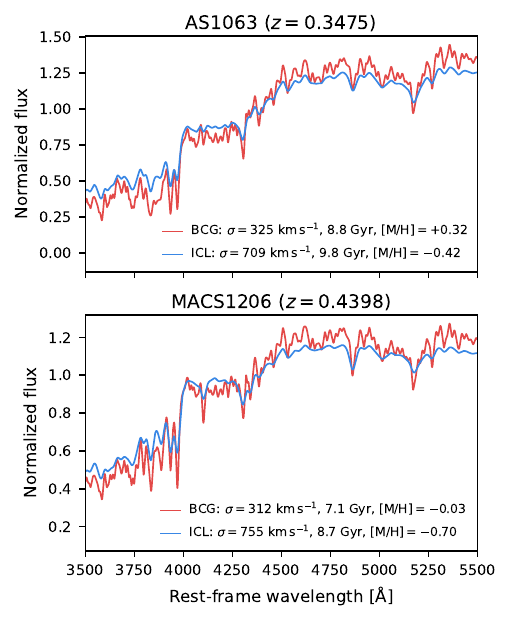}
    \caption{
    Optimal templates of the BCG (red) and ICL (blue) for AS1063 (top) and MACS1206 (bottom), derived through the iterative procedure described in \textsection~\ref{subsec:templates}.
    Each template is broadened by the best-fit velocity dispersion of the corresponding component and normalized by its median over the rest-frame $3500$--$5500~\text{\AA}$ range.
    The legend lists the velocity dispersion, light-weighted age, and metallicity of each template (see also Table~\ref{tab:templates}).
    In both clusters, the ICL shows significantly broader absorption features than the BCG and a lower metallicity (at the $\mytilde2\sigma$ level).
    }
    \label{fig:templates}
\end{figure}

To build a template for each component, we extracted two spectra from the sky-subtracted cube: a core spectrum within a circle of $r\leq5$~kpc, and an outskirt spectrum within the elliptical region of $80\leq\sma\leq120$~kpc (AS1063) or $70\leq\sma\leq110$~kpc (MACS1206).
Both regions are centered on the BCG, and the outskirt region follows the elliptical geometry of \textsection~\ref{subsec:sky}.
We fitted these spectra with the penalized pixel-fitting method (\texttt{pPXF}; \citealt{Cappellari2004, Cappellari2017, Cappellari2023}) using the E-MILES single stellar population (SSP) models \citep{Vazdekis2016} based on the BaSTI isochrones \citep{Pietrinferni2004, Pietrinferni2006}.%
\footnote{We included a sixth-order multiplicative Legendre polynomial (\texttt{mdegree}~$=6$) to absorb residual flux-calibration errors in the continuum shape, and no additive polynomial (\texttt{degree}~$=-1$) so as not to alter the depths of the absorption lines.}
We imposed a prior that limits the SSP ages to be younger than the age of the universe at each cluster redshift ($9.8$ and $9.0$~Gyr for AS1063 and MACS1206, respectively).
We restricted the fit to the rest-frame $3500$--$5500~\text{\AA}$ range for both clusters. The blue end is set by the blue limit of the MUSE wavelength coverage at the redshift of AS1063, and the red end by the increasingly strong residuals of the sky emission lines at longer wavelengths.
Because both spectra contain light from both components, we determined the templates iteratively.
\begin{enumerate}
    \item We fitted the outskirt spectrum with a single stellar component to obtain the initial ICL template.
    \item We fitted the core spectrum with two components: the BCG component, modeled with the full SSP library, and the ICL component, fixed to the current ICL template. This step provides the BCG template and the light weight of the ICL in the core spectrum ($w_{\mathrm{ICL}}$).
    \item We fitted the outskirt spectrum in the same manner, with the ICL component modeled with the full SSP library and the BCG component fixed to the BCG template from step 2. This step updates the ICL template and provides the light weight of the BCG in the outskirt spectrum ($w_{\mathrm{BCG}}$).
    \item We repeated steps 2 and 3 until both $w_{\mathrm{ICL}}$ and $w_{\mathrm{BCG}}$ converged, with changes smaller than 0.005 between successive iterations.
\end{enumerate}
The kinematic parameters in each step were sampled with the nested sampling algorithm \texttt{MultiNest} \citep{Feroz2008, Feroz2009, Feroz2019}.
The best-fit weighted combinations of the SSPs at convergence define the optimal BCG and ICL templates, which are shown in Figure~\ref{fig:templates} and summarized in Table~\ref{tab:templates}.
These templates were kept fixed in the subsequent bin-by-bin fits.

Both templates are dominated by old stellar populations, with light-weighted ages of $7$--$10$~Gyr (Table~\ref{tab:templates}).
The ages of the BCG and ICL agree within the uncertainties, and those of the ICL are bounded by the age prior, so that we cannot detect an age difference between the two components.
The ICL is more metal-poor than the BCG by $0.74^{+0.49}_{-0.44}$~dex in AS1063 and $0.67^{+0.37}_{-0.66}$~dex in MACS1206, corresponding to a combined statistical significance of $\mytilde2\sigma$.
This difference is in the sense expected if the ICL is built mainly from stars stripped from low-mass satellites and from the outskirts of massive ones (\textsection~\ref{sec:intro}), and is consistent with the negative color and metallicity gradients observed in BCG+ICL systems \citep[e.g.,][]{Montes2014, DeMaio2018}.
These are light-weighted values from models with a fixed abundance pattern, and their uncertainties do not include the systematic uncertainties of the stellar population models (\textsection~\ref{subsec:b26}).

The E-MILES templates (FWHM $=5.0$~\AA) are broader than the MUSE line-spread function at the rest-frame wavelengths of the clusters, and the fitted dispersions $\sigma_{\mathrm{fit}}$ therefore underestimate the intrinsic ones.
The intrinsic dispersion is $\sigma=(\sigma_{\mathrm{fit}}^{2}+\Delta\sigma^{2})^{1/2}$, where $\Delta\sigma^{2}=\sigma_{\mathrm{temp}}^{2}-\sigma_{\mathrm{MUSE}}^{2}$, and $\sigma_{\mathrm{temp}}$ and $\sigma_{\mathrm{MUSE}}$ are the template and instrumental resolutions in velocity units at the center of the fitting range ($4500$~\AA~rest frame), which gives $\Delta\sigma=130.0$ and $131.5~\mathrm{km~s^{-1}}$ for AS1063 and MACS1206, respectively.
This correction raises the dispersions of the BCG by $7$--$10\%$ and those of the ICL by only $\mytilde2\%$, and does not affect the kinematic contrast between the two components.
Throughout this paper, including Tables~\ref{tab:templates} and \ref{tab:priors} and Figure~\ref{fig:kinematics}, we quote the uncorrected $\sigma_{\mathrm{fit}}$.

\begin{table*}
\setlength{\tabcolsep}{4pt}
\caption{Properties of the Optimal BCG and ICL Templates \label{tab:templates}}
\begin{center}
\begin{tabular}{llcccc}
\hline\hline
Cluster & Comp. & $V$ & $\sigma$ & Age & [M/H] \\
 & & (km~s$^{-1}$) & (km~s$^{-1}$) & (Gyr) & \\
\hline
AS1063 & BCG & $-14.1^{+5.0}_{-4.7}$ & $324.6^{+11.5}_{-10.9}$ & $8.8^{+0.8}_{-2.5}$ & $+0.32^{+0.07}_{-0.16}$ \\
 & ICL & $-79.2\pm53.0$ & $708.8\pm90.1$ & $9.8^{+0.0}_{-2.2}$ & $-0.42^{+0.41}_{-0.48}$ \\
\hline
MACS1206 & BCG & $7.9^{+8.8}_{-7.9}$ & $311.9^{+10.7}_{-10.1}$ & $7.1^{+1.5}_{-2.1}$ & $-0.03^{+0.33}_{-0.66}$ \\
 & ICL & $-169.2^{+129.7}_{-135.8}$ & $755.3^{+159.5}_{-77.8}$ & $8.7^{+0.3}_{-3.3}$ & $-0.70^{+0.06}_{-0.17}$ \\
\hline
\end{tabular}
\end{center}
\tablecomments{Velocities are relative to the cluster redshift (\textsection~\ref{subsec:templates}). Ages and metallicities are light-weighted averages of the SSP weights of each template. The age posteriors of the ICL are truncated at the age of the universe at the cluster redshift (\textsection~\ref{subsec:templates}).}
\end{table*}

\subsection{Spatial Binning} \label{subsec:binning}
Because the ICL signal in individual spaxels is hardly distinguishable from the sky, we stacked the spectra spatially in two ways: elliptical annuli, which provide our primary measurements, and adaptive Voronoi bins, which test the azimuthal dependence of the annulus measurements.

We extracted the spectra of 10 elliptical annuli, logarithmically spaced between $\sma=1$ and 200~kpc, using the same geometry as in \textsection~\ref{subsec:sky}.
The annuli sacrifice azimuthal information but reach a very high $S/N$ in each bin, which allows the kinematics to be fitted with wide, flat priors (\textsection~\ref{subsec:kinematics}).

We also constructed adaptive Voronoi bins, which sample the profiles at finer radial steps.
For this binning, we first constructed an $S/N$ map from the sky-subtracted, masked data cube in the rest-frame $4800$--$5100~\text{\AA}$ window, where the stellar continuum is relatively bright while the contamination from sky emission lines remains moderate.
The noise of each spaxel includes three terms: the propagated variance, the empirical scatter of the flux within the window, and the uncertainty of the subtracted sky spectrum (\textsection~\ref{subsec:sky}).
We adopted the larger of the first two terms and added the sky term in quadrature.
We then applied the Voronoi tessellation \citep{Cappellari2003} to this $S/N$ map, which achieves a high spatial resolution near the bright core and progressively coarser bins toward the faint outskirts.
For AS1063, we adopted a target $S/N$ of 600, resulting in 54 bins.
Because the MACS1206 data are shallower than those of AS1063, the same target $S/N$ would leave too few bins to trace the radial variation.
Instead, we adopted a target $S/N$ of 350 within $\sma=30$~kpc, resulting in 22 bins, and a lower target $S/N$ of 160 beyond this radius, resulting in 39 bins.
Throughout this paper, we present the results from both binning schemes together, so that any systematic offset between the two would be directly visible.

\subsection{Kinematic Decomposition of the BCG and ICL} \label{subsec:kinematics}
For each bin, we fitted the spectrum with the two fixed templates, over the same wavelength range and with the same polynomial settings as in \textsection~\ref{subsec:templates}.
Spectral pixels affected by sky-subtraction residuals were identified as local excesses in the noise spectrum and excluded from the fit.
If the reduced $\chi^{2}_{\nu}$ of a preliminary fit exceeded unity, we rescaled the noise spectrum of the bin by $\sqrt{\chi^{2}_{\nu}}$.
The four kinematic parameters ($V_{\mathrm{BCG}}$, $\sigma_{\mathrm{BCG}}$, $V_{\mathrm{ICL}}$, $\sigma_{\mathrm{ICL}}$), which enter the model nonlinearly, were sampled with \texttt{PyMultiNest} \citep{Buchner2014}.
At each likelihood evaluation, the two template weights ($w_{\mathrm{BCG}}$, $w_{\mathrm{ICL}}$), which enter linearly, were obtained directly by non-negative least squares within \texttt{pPXF}, so that only the kinematics were explored by the sampler.

For the elliptical annuli, whose $S/N$ is high, we adopted flat priors on the four kinematic parameters, spanning $\pm6$ widths of the template posteriors for the BCG and $\pm3$ widths for the ICL (last column of Table~\ref{tab:priors}).
These ranges are much wider than the posteriors of the BCG parameters at all radii and of the ICL parameters at intermediate radii; only in the innermost and outermost annuli are the ICL kinematics prior-limited.
For the Voronoi bins, whose $S/N$ is much lower than that of the annuli, the fits require some regularization, and we instead adopted Gaussian priors centered on the best-fit template kinematics (\textsection~\ref{subsec:templates}), with widths equal to the mean of their asymmetric posterior uncertainties and truncated at $\pm3$ widths for the BCG and $\pm1.5$ widths for the ICL.

We excluded Voronoi bins whose best-fit multiplicative polynomial shows a peak-to-trough amplitude of $\geq50\%$ over the fitting range.
Because residual flux-calibration errors are only a few percent (\textsection~\ref{subsec:reduction}), such an amplitude indicates that the polynomial is compensating for a failed decomposition or for residual sky features.
This criterion excludes 4 of the 54 bins in AS1063 and 18 of the 61 bins in MACS1206, most of which are located beyond $\sma=120$~kpc, where the light is faint and heavily masked.
We also excluded the bins whose centroids lie in the sky region ($\sma>200$~kpc; \textsection~\ref{subsec:sky}), which removes a further 4 bins in AS1063 and 1 bin in MACS1206.
In total, 46 and 42 bins are used for AS1063 and MACS1206, respectively.

In some bins, the fit assigned zero weight to one of the two templates, i.e., the posterior median of $w_{\mathrm{ICL}}/(w_{\mathrm{BCG}}+w_{\mathrm{ICL}})$ is exactly 0 or 1.
Because the kinematics of a component with zero weight are not constrained by the data but only by the prior, we do not report the kinematics of that component in such bins, for both the Voronoi bins and the elliptical annuli.

\begin{table*}
\setlength{\tabcolsep}{6pt}
\caption{Priors on the Kinematic Parameters \label{tab:priors}}
\begin{center}
\begin{tabular}{llcccc}
\hline\hline
Cluster & Parameter & $\mu$ & $s$ & Gaussian Range (Voronoi Bins) & Flat Range (Elliptical Annuli) \\
 & & (km~s$^{-1}$) & (km~s$^{-1}$) & (km~s$^{-1}$) & (km~s$^{-1}$) \\
\hline
AS1063 & $V_{\mathrm{BCG}}$ & $-14.1$ & 4.8 & $[-28.7,~0.4]$ & $[-42.9,~14.7]$ \\
 & $\sigma_{\mathrm{BCG}}$ & 324.6 & 11.2 & $[291.1,~358.2]$ & $[257.4,~391.8]$ \\
 & $V_{\mathrm{ICL}}$ & $-79.2$ & 53.0 & $[-158.8,~0.3]$ & $[-238.2,~79.8]$ \\
 & $\sigma_{\mathrm{ICL}}$ & 708.8 & 90.1 & $[573.7,~843.9]$ & $[438.5,~979.1]$ \\
\hline
MACS1206 & $V_{\mathrm{BCG}}$ & 7.9 & 8.4 & $[-17.2,~33.0]$ & $[-42.5,~58.3]$ \\
 & $\sigma_{\mathrm{BCG}}$ & 311.9 & 10.4 & $[280.7,~343.1]$ & $[249.5,~374.3]$ \\
 & $V_{\mathrm{ICL}}$ & $-169.2$ & 132.7 & $[-368.3,~29.9]$ & $[-567.3,~228.9]$ \\
 & $\sigma_{\mathrm{ICL}}$ & 755.3 & 118.6 & $[577.3,~933.2]$ & $[399.5,~1111.1]$ \\
\hline
\end{tabular}
\end{center}
\tablecomments{For the Voronoi bins, each prior is a Gaussian with mean $\mu$ and standard deviation $s$, taken from the best-fit value and posterior width of the BCG or ICL spectrum fit (\textsection~\ref{subsec:templates}), truncated at $\pm3s$ for the BCG and $\pm1.5s$ for the ICL (Gaussian Range). For the elliptical annuli, each prior is flat over $\pm6s$ for the BCG and $\pm3s$ for the ICL (Flat Range). Velocities are relative to the cluster redshift (\textsection~\ref{subsec:templates}).}
\end{table*}

\section{Results} \label{sec:result}

\subsection{Radial Kinematic Profiles} \label{subsec:kinprofile}
\begin{figure*}
    \centering
    \includegraphics[width=\textwidth]{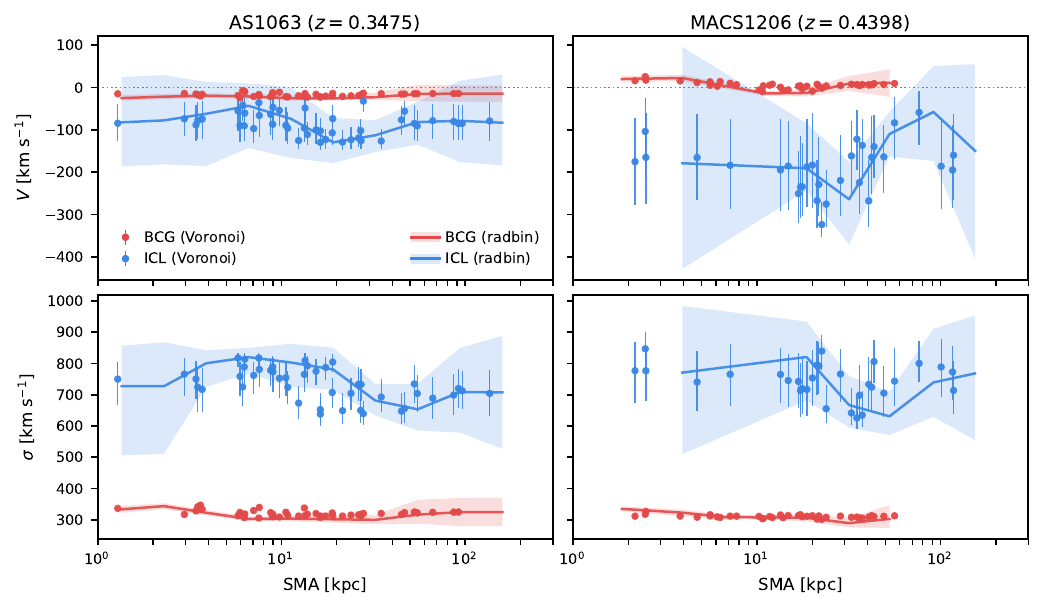}
    \caption{Radial profiles of the line-of-sight velocity ($V$; left) and velocity dispersion ($\sigma$; right) of the BCG (red) and ICL (blue) components for AS1063 (top) and MACS1206 (bottom).
    Points show the Voronoi bins, and the solid lines and shaded regions show the median and the 16th--84th percentile range of the posteriors in the elliptical annuli.
    Error bars of the Voronoi bins indicate the 16th--84th percentile range of the posteriors.
    Bins rejected by the quality control and components with zero template weight are not shown.
    \label{fig:kinematics}}
\end{figure*}

Figure~\ref{fig:kinematics} shows the kinematic profiles; the values quoted below are medians over the elliptical annuli.

In both clusters, the two components are clearly distinct in velocity and velocity dispersion.
In AS1063, the median velocities of the BCG and ICL are $-22\pm2$ and $-97\pm22~\mathrm{km~s^{-1}}$, respectively, so that the ICL is blueshifted relative to the BCG by $75\pm22~\mathrm{km~s^{-1}}$.
In MACS1206, they are $+2\pm6$ and $-146\pm38~\mathrm{km~s^{-1}}$, corresponding to a larger offset of $148\pm38~\mathrm{km~s^{-1}}$.
Such an offset may reflect a relative line-of-sight motion between the BCG and the cluster-scale halo traced by the ICL; in simulated cluster mergers, the BCG remains offset from the DM halo center for several Gyr after pericenter \citep{Lee2026}.
Because DM self-interaction produces cored halos \citep{Lee2026}, in which a displaced BCG experiences a weaker restoring force, the BCG--ICL velocity offset measured for a larger sample could provide a line-of-sight probe of DM physics that is complementary to constraints from cluster mergers \citep[e.g.,][]{Jee2026}.

The median velocity dispersions of the BCG and ICL are $316\pm5$ and $753\pm25~\mathrm{km~s^{-1}}$ in AS1063 and $313\pm5$ and $695^{+47}_{-45}~\mathrm{km~s^{-1}}$ in MACS1206, respectively.
The ICL is dynamically hotter than the BCG by a factor of $2.39\pm0.09$ in AS1063 and $2.22\pm0.15$ in MACS1206.
This is consistent with the expectation that the BCG stars are bound to the potential of the BCG itself, whereas the ICL stars trace the potential of the cluster-scale halo \citep{Dolag2010}.
Relative to the member galaxies, with dispersions of $1380$ (AS1063; \citealt{Mercurio2021}) and $1087~\mathrm{km~s^{-1}}$ (MACS1206; \citealt{Biviano2013}), the ICL dispersions are $\mytilde0.55$ and $\mytilde0.64$ times as large, although the galaxy dispersion of AS1063 may be inflated by its bimodal velocity distribution (\textsection~\ref{subsec:targets}).
In equilibrium, the dispersion of a tracer depends on the mass enclosed within its radius and on the slope of its density profile; the ICL is confined to the inner $\mytilde150$~kpc and declines much more steeply than the galaxy distribution, both of which lower its dispersion relative to that of the galaxies.

Within the uncertainties, neither component shows a significant radial trend in $V$ or $\sigma$ in either cluster.
The Voronoi bins agree with the annuli within the uncertainties at all radii, indicating that the kinematics of both components do not vary significantly with azimuth at the precision of our data.

\subsection{Surface Brightness Decomposition} \label{subsec:sbdecomp}
\begin{figure*}
    \centering
    \includegraphics[width = \textwidth]{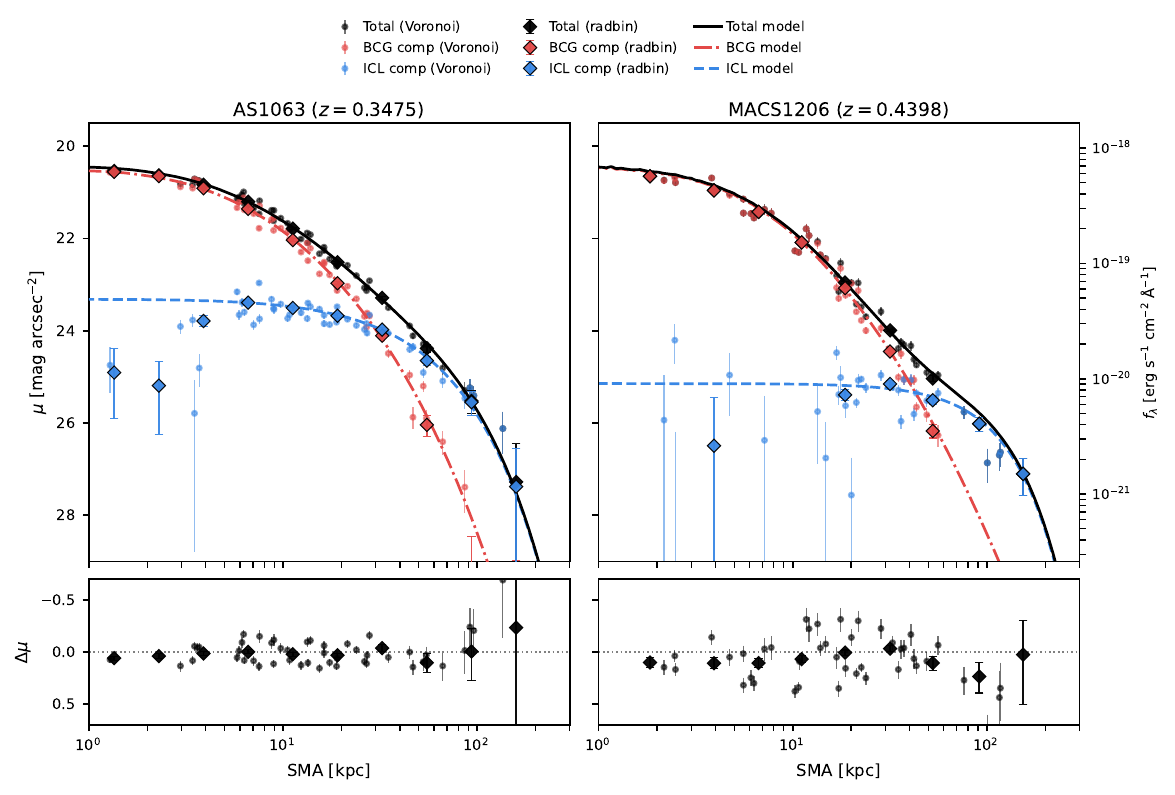}
    \caption{Surface brightness profiles of AS1063 (left) and MACS1206 (right).
    The surface brightness is measured over the rest-frame $3500$--$5500~\text{\AA}$ range, and the total light (black) is decomposed into the BCG (red) and ICL (blue) components using the light weights of the kinematic decomposition.
    Small circles and diamonds show the Voronoi bins and the elliptical annuli, respectively, and the error bars include the uncertainty of the sky level.
    The dash-dotted and dashed lines show the PSF-convolved single S\'ersic models of the BCG and ICL, respectively, and the solid line shows their sum.
    The bottom panels show the residuals of the total surface brightness from the summed model.
    The right-hand axis gives the corresponding flux density per spaxel for MACS1206; the conversion depends on the cluster redshift and is therefore not shown for AS1063.
    \label{fig:sfbprofile}}
\end{figure*}

Figure~\ref{fig:sfbprofile} shows the surface brightness (SB) profiles of the BCG and ICL obtained by multiplying the SB of each bin by the light fractions $w_{\mathrm{BCG}}/(w_{\mathrm{BCG}}+w_{\mathrm{ICL}})$ and $w_{\mathrm{ICL}}/(w_{\mathrm{BCG}}+w_{\mathrm{ICL}})$.
This decomposition assumes that the stellar populations of each component do not vary with radius.
The BCG dominates the central region, while the ICL takes over beyond $\sma\approx30$~kpc in AS1063 and $\approx43$~kpc in MACS1206, and is traced to $\sma\approx160$~kpc in both clusters.
The Voronoi bins and the elliptical annuli agree with each other over the entire radial range, consistent with the weak azimuthal dependence found in the kinematic profiles (\textsection~\ref{subsec:kinprofile}).

To characterize the structures of the two components, we fitted a \sersic model \citep{Sersic1963} to the BCG and ICL profiles from the elliptical annuli (\textsection~\ref{subsec:binning}),
\begin{equation}
    I(R) = I_{e} \exp\left[-b_{n}\left(\left(\frac{R}{R_{e}}\right)^{1/n}-1\right)\right],
    \label{eq:sersic}
\end{equation}
where $R_{e}$ is the effective radius, $I_{e}$ is the intensity at $R_{e}$, $n$ is the \sersic index, and $b_{n}$ is a constant determined solely by $n$.
Because the PSF redistributes light from the bright core to larger radii, we fitted the profiles by forward modeling, convolving each \sersic model with the PSF before comparing it with the observed profiles, so that the fitted parameters describe the intrinsic, PSF-corrected profiles of the two components.
For AS1063, we combined the PSFs of the AO-mode and standard-mode exposures weighted by their exposure times, where the former was reconstructed from the AO telemetry with \texttt{muse\_psfr} \citep{Fusco2020} and the latter was estimated from the seeing recorded during the observations. The combined PSF is described by a Moffat profile with $\mathrm{FWHM}=0.43^{\prime\prime}$ and $\beta=1.60$.
For the MACS1206 data, where no AO telemetry is available, we fitted a two-dimensional Moffat profile to a field star in the collapsed image and obtained $\mathrm{FWHM}=0.61^{\prime\prime}$ and $\beta=1.69$.
We constructed the PSF kernel over the full extent of the data and performed the fits with \texttt{dynesty} \citep{Speagle2020}; Table~\ref{tab:sersic} lists the best-fit parameters.

The kinematically decomposed BCG and ICL are each well described by a single \sersic profile, and the sum of the two models successfully reproduces the total SB profiles of both clusters within $\mytilde0.2$~mag out to $\sma\approx100$~kpc.
The BCG of MACS1206 is smaller than that of AS1063, $R_{e}=12.0^{+1.2}_{-0.6}$~kpc versus $17.7^{+1.3}_{-0.9}$~kpc, and more centrally concentrated, with $n=2.88^{+0.56}_{-0.21}$ compared with $1.87^{+0.18}_{-0.13}$ in AS1063.
The ICL components are much more extended, with $R_{e}=64.7^{+28.2}_{-5.0}$~kpc and $n=0.90^{+0.31}_{-0.08}$ in AS1063 and $R_{e}=95.9^{+32.8}_{-5.6}$~kpc and $n=0.56^{+0.28}_{-0.11}$ in MACS1206.
Within $\sma=200$~kpc, the ICL defined by our kinematic decomposition (hereafter the kinematic ICL) accounts for $60^{+6}_{-2}\%$ of the total light in AS1063 and $51^{+4}_{-2}\%$ in MACS1206, in the rest-frame $3500$--$5500~\text{\AA}$ band.
We compare this kinematic decomposition with multi-component profile fitting in \textsection~\ref{subsec:multisersic}.

\begin{deluxetable}{llcc}
\tablecaption{PSF-corrected \sersic Parameters \label{tab:sersic}}
\tablehead{
\colhead{Cluster} & \colhead{Component} & \colhead{$R_{e}$} & \colhead{$n$} \\
\colhead{} & \colhead{} & \colhead{(kpc)} & \colhead{}
}
\startdata
AS1063 & BCG & $17.68^{+1.29}_{-0.94}$ & $1.87^{+0.18}_{-0.13}$ \\
 & ICL & $64.73^{+28.17}_{-5.03}$ & $0.90^{+0.31}_{-0.08}$ \\
\hline
MACS1206 & BCG & $12.02^{+1.19}_{-0.61}$ & $2.88^{+0.56}_{-0.21}$ \\
 & ICL & $95.93^{+32.79}_{-5.58}$ & $0.56^{+0.28}_{-0.11}$ \\
\enddata
\tablecomments{Best-fit parameters of the BCG and ICL profiles from the elliptical annuli, with uncertainties given by the 16th and 84th percentiles of the posterior distributions. Fits to the Voronoi-bin profiles give consistent results.}
\end{deluxetable}

\section{Discussion} \label{sec:discussion}
\subsection{Kinematic Decomposition versus Conventional BCG--ICL Separations} \label{subsec:boundary}
We use the kinematic decomposition to test how much of the kinematic ICL each conventional definition (an SB threshold, a fixed radius, or multi-component fitting of the light profile) captures, and where the boundary between the two components should lie. For the first two, we applied each boundary to the PSF-convolved best-fit \sersic models of the BCG and ICL (\textsection~\ref{subsec:sbdecomp}), which reproduce the observed light, and computed the fraction of the kinematic ICL light enclosed within it and thus assigned to the BCG (hereafter the ICL loss).

We first converted the SB threshold of $\mu_{B} = 25 ~\mathrm{mag~arcsec^{-2}}$ to our rest-frame $3500$--$5500~\text{\AA}$ band and to the redshift of each cluster, using the synthetic photometry of the ICL spectrum extracted from the outskirt region and correcting for the cosmological dimming.
The converted thresholds are $25.91$ and $26.41~\mathrm{mag~arcsec^{-2}}$ for AS1063 and MACS1206, which are reached at $\sma = 106.0^{+17.7}_{-4.5}$ and $123.6^{+13.0}_{-4.0}$~kpc, respectively.
At these radii, the ICL already dominates the local light, with ICL fractions of $92^{+4}_{-3}\%$ (AS1063) and $93^{+3}_{-3}\%$ (MACS1206).
However, since these radii lie beyond the effective radii of the kinematic ICL ($\mytilde65$--$95$~kpc; Table~\ref{tab:sersic}), two-thirds to three-quarters of the kinematic ICL light is enclosed within these radii and attributed to the BCG, with ICL losses of $77^{+3}_{-12}\%$ in AS1063 and $66^{+7}_{-16}\%$ in MACS1206.

For the radius-based definition, a fixed aperture of 50~kpc yields smaller ICL losses of $36^{+4}_{-12}\%$ (AS1063) and $17^{+1}_{-5}\%$ (MACS1206).
However, the same radius corresponds to different positions relative to the transition between the two components. The ICL contributes $73^{+2}_{-2}\%$ of the local light at 50~kpc in AS1063, but only $60^{+2}_{-3}\%$ in MACS1206, where the ICL begins to dominate at a larger radius ($42.9^{+1.7}_{-1.0}$~kpc versus $29.1^{+1.2}_{-0.6}$~kpc in AS1063).
The 50~kpc aperture thus lies at $1.72^{+0.04}_{-0.07}$ and $1.17^{+0.03}_{-0.05}$ times the transition radius in AS1063 and MACS1206, respectively, 
and an aperture scaled to this radius, rather than a fixed physical one, would misclassify comparable fractions of the kinematic ICL light in the two clusters.

\citet{Contini2026} proposed defining the boundary between the BCG and ICL as the three-dimensional aperture radius $r_{\mathrm{cut}}$ at which the BCG mass outside the aperture equals the ICL mass inside it.
They further showed that $r_{\mathrm{cut}}$ scales with the virial radius, as $r_{\mathrm{cut}}/R_{\mathrm{vir}}=0.045-0.007\log(M_{\mathrm{halo}}/10^{14}~M_{\odot})$.\footnote{Their simulated halos span $\log M_{\mathrm{halo}}/M_{\odot}\lesssim15$, so both the relation and its scatter are extrapolated to our clusters. For the scatter, we adopted $\pm0.01$ in $r_{\mathrm{cut}}/R_{\mathrm{vir}}$, the 16th--84th percentile range of their halos at the high-mass end of their sample (their Figure~2).}
Since we did not derive the stellar mass profiles, we applied the condition to the light profiles instead of the mass profiles, assuming a common mass-to-light ratio for the BCG and ICL.
Applied to the projected, PSF-corrected profiles, which are directly measured from the data, the condition gives $r_{\mathrm{cut}}=30.9^{+1.8}_{-1.2}$~kpc (AS1063) and $41.5^{+4.5}_{-2.3}$~kpc (MACS1206).
These are comparable to, although somewhat larger than, the radii where the ICL accounts for half of the surface brightness in the same PSF-corrected models, which are $28.1$ and $38.4$~kpc, respectively.
The balance condition of \citet{Contini2026} thus places the boundary close to the radius where the BCG and ICL contribute equally to the surface brightness, suggesting that the definition itself provides a reasonable boundary between the two components.

We also calculated $r_{\mathrm{cut}}$ from the scaling relation, adopting the lensing masses of AS1063 \citep{Gruen2013} and MACS1206 \citep{Umetsu2012},%
\footnote{\citet{Umetsu2012} reported the virial mass of MACS1206, $M_{\mathrm{vir}}=(1.1\pm0.2\pm0.1)\times10^{15}~h^{-1}M_{\odot}$, which we converted to $M_{\odot}$ with our value of $h$.
For AS1063, we converted the mass and concentration of \citet{Gruen2013}, $M_{200m}=33.1^{+9.6}_{-6.8}\times10^{14}~M_{\odot}$ and $c_{200m}=2.6^{+1.5}_{-1.0}$, into the virial mass assuming an NFW profile \citep{Navarro1997}.
For both clusters, we computed the virial radius with the virial overdensity of \citet{Bryan1998} in our cosmology.
The uncertainties in the masses, the concentration, and the scatter of the scaling relation were propagated with Monte Carlo sampling.} which yields $r_{\mathrm{cut}}=107.6^{+32.0}_{-30.4}$~kpc for AS1063 and $87.2^{+24.7}_{-24.0}$~kpc for MACS1206.
Our $r_{\mathrm{cut}}$ values are smaller than these predictions by a factor of $3.5^{+0.9}_{-1.0}$ in AS1063 and $2.1^{+0.6}_{-0.6}$ in MACS1206.
Since the prediction is based on the intrinsic three-dimensional distributions of the BCG and ICL, we deprojected our best-fit profiles into three-dimensional luminosity densities using the Abel inversion \citep[e.g.,][]{BinneyTremaine2008} and derived $r_{\mathrm{cut}}$ with spherical apertures.
The deprojection increases $r_{\mathrm{cut}}$ to $44.0^{+3.6}_{-2.0}$~kpc for AS1063 and $60.6^{+7.8}_{-4.1}$~kpc for MACS1206.
For MACS1206, the deprojected $r_{\mathrm{cut}}$ is consistent with the prediction within the uncertainties.
For AS1063, however, the deprojected $r_{\mathrm{cut}}$ is only $\mytilde 41\%$ of the prediction.

Multi-component fitting separates the two components without an explicit boundary, by fitting the total light profile with multiple \sersic components \citep[e.g.,][]{Gonzalez2005, Zibetti2005, Kluge2020, Joo2023}. \label{subsec:multisersic}
We fitted two \sersic components simultaneously to the total SB profiles, with the same PSF, fitting procedure, and prior ranges as for the individual BCG and ICL fits (\textsection~\ref{subsec:sbdecomp}).
The fit reproduces the total profiles as well as the sum of the individual BCG and ICL models does.
Within $\sma=200$~kpc, the outer component accounts for $86^{+8}_{-11}\%$ and $84^{+7}_{-12}\%$ of the total light in AS1063 and MACS1206, respectively, compared with $60^{+6}_{-2}\%$ and $51^{+4}_{-2}\%$ for the kinematic ICL.
Light can be exchanged between the two components with little change in the total profile, and the \sersic index of the outer component is degenerate with the size and concentration of the inner component.
The result therefore depends on the adopted priors. Limiting the outer \sersic index to $n_{2}\leq2$, as found for the kinematic ICL (Table~\ref{tab:sersic}) and for the outer components of most clusters in deep HST imaging \citep{Joo2023}, reduces the outer fractions to $71^{+15}_{-13}\%$ and $54^{+17}_{-14}\%$, respectively, closer to the kinematic ICL.
Such a prior, however, is not universal, since some clusters show outer components with $n\gtrsim3$ \citep[e.g.,][]{Gonzalez2005}.
Given this degeneracy and the depth and spatial resolution of the MUSE data, the BCG and ICL cannot be reliably separated by a simultaneous fit of two \sersic components in our clusters.
Whether deeper and higher-resolution imaging from HST and JWST can break this degeneracy will be addressed in future work.

\subsection{Rising Velocity Dispersion Profiles of BCGs} \label{subsec:rising}
The outward rise in BCG velocity dispersion (\textsection~\ref{sec:intro}) has been used to trace the cluster mass distribution with the BCG stars as a single tracer \citep[e.g.,][]{Kelson2002, Newman2013}, and in a few nearby cD galaxies it has been attributed to a halo bound to the cluster rather than to the galaxy \citep{Bender2015, Barbosa2018}.
Our decomposition supports the latter interpretation and shows that the changing mixture of the two components can account for the rise.

In both clusters, the dispersion of each component shows no significant variation with radius (\textsection~\ref{subsec:kinprofile}), whereas the light fraction of the ICL increases outward (\textsection~\ref{subsec:sbdecomp}).
A single-component fit to the combined light would yield a dispersion that rises from that of the BCG ($\mytilde315~\mathrm{km~s^{-1}}$) toward that of the ICL ($\mytilde700$--$750~\mathrm{km~s^{-1}}$) as the ICL contribution grows \citep[see][for AS1063]{Bellhouse2026}.
\citet{Hilker2018} reached a similar conclusion for NGC~3311.
In this picture, the radius at which the dispersion begins to rise marks the BCG--ICL transition and need not indicate a change in the potential. Dynamical models that treat the BCG as a single tracer then attribute the ICL kinematics to the BCG, which can bias the inferred mass profile where the ICL contributes appreciable light.

\subsection{Comparison with \citet{Bellhouse2026}} \label{subsec:b26}
\citet{Bellhouse2026} decomposed the BCG and ICL of AS1063 with a method similar to ours, fitting the spectrum of each spatial bin with two fixed templates using \texttt{pPXF}. Their data consist of two standard-mode pointings with a total integration time of $7.9$~hr, both of which are included in our data set (\textsection~\ref{subsec:obs}).
The kinematics of the two analyses agree well.
Their template dispersions, $\sigma_{\mathrm{BCG}}=357$ and $\sigma_{\mathrm{ICL}}=791~\mathrm{km~s^{-1}}$, are consistent with ours (Table~\ref{tab:templates}) within the uncertainty of our ICL dispersion, once the resolution correction (\textsection~\ref{subsec:templates}) is taken into account.
The dispersion profiles are nearly flat in both analyses, and the ICL-to-BCG dispersion ratio is $2.39\pm0.09$ in ours and $2.1$ in theirs.
The ICL is blueshifted relative to the BCG by $75\pm22$ and $86\pm18~\mathrm{km~s^{-1}}$, respectively.
The transition between the two components lies at $\mytilde30$~kpc in our analysis and at $\mytilde35$~kpc in theirs.
\citet{Bellhouse2026} also found that the photometric decomposition of \citet{Morishita2017} assigns a substantial part of the kinematic ICL to the BCG, as we found for the SB threshold and fixed apertures (\textsection~\ref{subsec:boundary}).

In contrast, the light distributions and stellar populations differ.
The ICL of \citet{Bellhouse2026} has $R_{e}=210\pm20$~kpc and $n=1.5\pm0.1$, about three times the $R_{e}$ of ours ($R_{e}=64.7^{+28.2}_{-5.0}$~kpc and $n=0.90^{+0.31}_{-0.08}$; Table~\ref{tab:sersic}).
Correspondingly, they find that the ICL emits $67\%$ of the BCG+ICL light within 100~kpc, compared with $\mytilde53\%$ from our models.
The metallicity difference between the BCG and ICL templates is $\mytilde0.2$~dex in their analysis ($[\mathrm{M/H}]=+0.26$ and $+0.05$), but $\mytilde0.7$~dex in ours ($+0.32^{+0.07}_{-0.16}$ and $-0.42^{+0.41}_{-0.48}$), although the two are consistent within our uncertainties.
Since the two data sets overlap, these differences are likely to arise mainly from the analysis choices, of which we identify three.

The first concerns the background, which both analyses estimate beyond $\sma\approx200$~kpc, although \citet{Bellhouse2026} detect ICL at these radii in the HST images.
They correct for the resulting oversubtraction by adding a constant of $0.03~\mu\mathrm{Jy~arcsec^{-2}}$ ($\mytilde27.7~\mathrm{mag~arcsec^{-2}}$ in F606W) to their ICL model before measuring its \sersic parameters. We applied no such correction, because the residual background in our data shows no radial trend beyond 200~kpc (\textsection~\ref{subsec:sky}).
A constant added to the ICL model flattens its outer profile and increases its $R_{e}$. Conversely, any ICL light remaining in our sky region would make our ICL more compact.

In the template construction, \citet{Bellhouse2026} remove the ICL contribution from their BCG template but not the BCG contribution from their ICL template, which they extract at $50$--$100$~kpc, closer to the BCG than our outskirt region.
From our \sersic models, the BCG contributes $\mytilde16\%$ of the light at $50$--$100$~kpc in AS1063, compared with $\mytilde9\%$ in our outskirt region ($80$--$120$~kpc), which our iterative procedure removes from the ICL template (\textsection~\ref{subsec:templates}).
BCG light left in the ICL template would raise its metallicity, reducing the contrast between the two templates, and would assign part of the BCG outskirts to the ICL.
\citet{Bellhouse2026} also assign all the light in bins fainter than $25~\mathrm{mag~arcsec^{-2}}$ in F606W to the ICL. We decompose the light into the two components out to $\sma=200$~kpc.

Finally, \citet{Bellhouse2026} fit the $\alpha$-element abundance as a free parameter.
The E-MILES models we used have a fixed abundance pattern, and because the metallicities of old populations are degenerate with $[\alpha/\mathrm{Fe}]$, this difference can change the metallicities of both templates.

These choices leave the kinematics of the two components largely unchanged but affect the light distribution and the stellar population of the ICL.
Because our data include their exposures, rerunning either analysis on the same data with the choices of the other would identify which of them drives the differences.

\section{Conclusions} \label{sec:conclusion}

In this work, we separated the BCG and ICL of AS1063 ($z=0.35$) and MACS1206 ($z=0.44$) by their stellar kinematics, using deep VLT/MUSE integral-field spectroscopy and two stellar templates derived iteratively from the BCG core and the ICL-dominated outskirts.
The resulting light profiles of the two components serve as a reference for photometric definitions of the ICL.

The BCG and ICL are kinematically distinct in both clusters out to $\sma\approx160$~kpc.
The ICL is dynamically hotter than the BCG by a factor of $2.39\pm0.09$ (AS1063) and $2.22\pm0.15$ (MACS1206) and is blueshifted relative to it by $75\pm22$ and $148\pm38~\mathrm{km~s^{-1}}$, respectively, consistent with the ICL stars tracing the cluster-scale potential rather than that of the BCG.
It is also more metal-poor than the BCG by $\mytilde0.7$~dex at a combined significance of $\mytilde2\sigma$, as expected if it is built mainly from stripped stars.
The BCG dominates the light within $\mytilde30$ (AS1063) and $\mytilde43$~kpc (MACS1206), and each component follows a single \sersic profile, with the ICL much more extended ($R_{e}\approx65$--$95$~kpc) and less centrally concentrated ($n\approx0.6$--$0.9$) than the BCG.
The ICL therefore forms a separate component that overlaps with the BCG over several tens of kpc.
Because each component has a nearly flat dispersion profile, the rising dispersion profiles observed in many BCGs can be explained by the radially increasing contribution of the ICL.

This overlap makes conventional photometric definitions misclassify a substantial and system-dependent fraction of the ICL.
The SB threshold of $\mu_{B}=25~\mathrm{mag~arcsec^{-2}}$ assigns two-thirds to three-quarters of the kinematic ICL to the BCG. A fixed 50~kpc aperture misclassifies less ($17$--$36\%$), but lies at different positions relative to the BCG--ICL transition in the two clusters.
A double \sersic model describes the total profiles well, but with our data it leaves the division of light between the two components poorly constrained, unless a prior such as $n_{2}\leq2$ is imposed.

Applying this approach to a larger sample of clusters spanning a range of halo masses and dynamical states would establish how the BCG--ICL boundary and the structures of the two components depend on the host cluster.
Such a sample would also allow us to measure the distribution of BCG--ICL velocity offsets, which traces the motion of BCGs within their host halos and is sensitive to the self-interaction of DM.
The ICL-free BCG kinematics, combined with strong-lensing mass models in a joint lensing and kinematic analysis, could also help constrain the inner DM profiles of clusters.

\begin{acknowledgments}
We thank Callum Bellhouse, whose presentation of a kinematic decomposition of the BCG and ICL in AS1063 with MUSE at the 2025 Sesto conference motivated this work, and Nina Hatch for helpful discussions and for kindly noting that the MUSE data of AS1063 are publicly available.
M.J.J. acknowledges support for the current research from the National Research Foundation (NRF) of Korea under the programs 2022R1A2C1003130 and RS-2026-25607498.
This work is based on observations collected at the European Southern Observatory under ESO programs 60.A-9345(A), 095.A-0653(A), 095.A-0181(A), 097.A-0269(A), 114.28L1.001, and 116.28VK.001.
This research is also based on observations made with the NASA/ESA Hubble Space Telescope obtained from the Space Telescope Science Institute, which is operated by the Association of Universities for Research in Astronomy, Inc., under NASA contract NAS 5-26555.
The HST data used in this work were obtained from the Mikulski Archive for Space Telescopes (MAST) at the Space Telescope Science Institute and can be accessed via \dataset[doi:10.17909/T9KK5N]{https://doi.org/10.17909/T9KK5N} (Frontier Fields) and \dataset[doi:10.17909/T90W2B]{https://doi.org/10.17909/T90W2B} (CLASH).
\end{acknowledgments}

\facilities{VLT:Yepun (MUSE), HST (ACS)}

\software{MUSE pipeline \citep{Weilbacher2020}, EsoRex, SExtractor \citep{Bertin1996}, photutils \citep{Bradley2024}, pPXF \citep{Cappellari2004, Cappellari2017, Cappellari2023}, MultiNest \citep{Feroz2008, Feroz2009, Feroz2019}, PyMultiNest \citep{Buchner2014}, muse\_psfr \citep{Fusco2020}, dynesty \citep{Speagle2020}}

\bibliography{reference}

@ARTICLE{Bellhouse2026,
       author = {{Bellhouse}, Callum and {Bamford}, Steven and {Hatch}, Nina and {Kolcu}, Tutku and {Richard}, Johan},
        title = "{Kinematically isolating intracluster light in AS1063: a distinct stellar population tracing the dark matter core}",
      journal = {arXiv e-prints},
         year = 2026,
        month = oct,
          eid = {arXiv:2610.28013},
        pages = {arXiv:2610.28013},
          doi = {10.48550/arXiv.2610.28013},
archivePrefix = {arXiv},
       eprint = {2610.28013},
 primaryClass = {astro-ph.GA}
}

@ARTICLE{Biviano2023,
       author = {{Biviano}, Andrea and {Pizzuti}, Lorenzo and {Mercurio}, Amata and {Sartoris}, Barbara and {Rosati}, Piero and {Ettori}, Stefano and {Girardi}, Marisa and {Grillo}, Claudio and {Caminha}, Gabriel B. and {Nonino}, Mario},
        title = "{CLASH-VLT: The Inner Slope of the MACS J1206.2-0847 Dark Matter Density Profile}",
      journal = {\apj},
         year = 2023,
        month = dec,
       volume = {958},
       number = {2},
          eid = {148},
        pages = {148},
          doi = {10.3847/1538-4357/acf832},
archivePrefix = {arXiv},
       eprint = {2307.06804},
 primaryClass = {astro-ph.CO},
       adsurl = {https://ui.adsabs.harvard.edu/abs/2023ApJ...958..148B}
}

@ARTICLE{Bohringer2004,
       author = {{B{\"o}hringer}, H. and {Schuecker}, P. and {Guzzo}, L. and {Collins}, C.~A. and {Voges}, W. and {Cruddace}, R.~G. and {Ortiz-Gil}, A. and {Chincarini}, G. and {De Grandi}, S. and {Edge}, A.~C. and {MacGillivray}, H.~T. and {Neumann}, D.~M. and {Schindler}, S. and {Shaver}, P.},
        title = "{The ROSAT-ESO Flux Limited X-ray (REFLEX) Galaxy cluster survey. V. The cluster catalogue}",
      journal = {\aap},
         year = 2004,
        month = oct,
       volume = {425},
        pages = {367-383},
          doi = {10.1051/0004-6361:20034484},
archivePrefix = {arXiv},
       eprint = {astro-ph/0405546},
 primaryClass = {astro-ph},
       adsurl = {https://ui.adsabs.harvard.edu/abs/2004A&A...425..367B}
}

@ARTICLE{Contini2026,
       author = {{Contini}, Emanuele and {Yi}, Sukyoung K.},
        title = "{On the Separation Between the Brightest Cluster Galaxy and the Intracluster Light}",
      journal = {arXiv e-prints},
         year = 2026,
        month = jul,
          eid = {arXiv:2607.17616},
        pages = {arXiv:2607.17616},
          doi = {10.48550/arXiv.2607.17616},
archivePrefix = {arXiv},
       eprint = {2607.17616},
 primaryClass = {astro-ph.GA},
       adsurl = {https://ui.adsabs.harvard.edu/abs/2026arXiv260717616C}
}

@ARTICLE{Umetsu2012,
       author = {{Umetsu}, Keiichi and {Medezinski}, Elinor and {Nonino}, Mario and {Merten}, Julian and {Zitrin}, Adi and {Molino}, Alberto and {Grillo}, Claudio and {Carrasco}, Mauricio and {Donahue}, Megan and {Mahdavi}, Andisheh and {Coe}, Dan and {Postman}, Marc and {Koekemoer}, Anton and {Czakon}, Nicole and {Sayers}, Jack and {Mroczkowski}, Tony and {Golwala}, Sunil and {Koch}, Patrick M. and {Lin}, Kai-Yang and {Molnar}, Sandor M. and {Rosati}, Piero and {Balestra}, Italo and {Mercurio}, Amata and {Scodeggio}, Marco and {Biviano}, Andrea and {Anguita}, Timo and {Infante}, Leopoldo and {Seidel}, Gregor and {Sendra}, Irene and {Jouvel}, Stephanie and {Host}, Ole and {Lemze}, Doron and {Broadhurst}, Tom and {Meneghetti}, Massimo and {Moustakas}, Leonidas and {Bartelmann}, Matthias and {Ben{\'\i}tez}, Narciso and {Bouwens}, Rychard and {Bradley}, Larry and {Ford}, Holland and {Jim{\'e}nez-Teja}, Yolanda and {Kelson}, Daniel and {Lahav}, Ofer and {Melchior}, Peter and {Moustakas}, John and {Ogaz}, Sara and {Seitz}, Stella and {Zheng}, Wei},
        title = "{CLASH: Mass Distribution in and around MACS J1206.2-0847 from a Full Cluster Lensing Analysis}",
      journal = {\apj},
         year = 2012,
        month = aug,
       volume = {755},
       number = {1},
          eid = {56},
        pages = {56},
          doi = {10.1088/0004-637X/755/1/56},
archivePrefix = {arXiv},
       eprint = {1204.3630},
 primaryClass = {astro-ph.CO},
       adsurl = {https://ui.adsabs.harvard.edu/abs/2012ApJ...755...56U}
}

@ARTICLE{Gruen2013,
       author = {{Gruen}, D. and {Brimioulle}, F. and {Seitz}, S. and {Lee}, C.-H. and {Young}, J. and {Koppenhoefer}, J. and {Eichner}, T. and {Riffeser}, A. and {Vikram}, V. and {Weidinger}, T. and {Zenteno}, A.},
        title = "{Weak lensing analysis of RXC J2248.7-4431}",
      journal = {\mnras},
         year = 2013,
        month = jun,
       volume = {432},
       number = {2},
        pages = {1455-1467},
          doi = {10.1093/mnras/stt566},
archivePrefix = {arXiv},
       eprint = {1304.0764},
 primaryClass = {astro-ph.CO},
       adsurl = {https://ui.adsabs.harvard.edu/abs/2013MNRAS.432.1455G}
}

@ARTICLE{Newman2013,
       author = {{Newman}, Andrew B. and {Treu}, Tommaso and {Ellis}, Richard S. and {Sand}, David J.},
        title = "{The Density Profiles of Massive, Relaxed Galaxy Clusters. II. Separating Luminous and Dark Matter in Cluster Cores}",
      journal = {\apj},
         year = 2013,
        month = mar,
       volume = {765},
       number = {1},
          eid = {25},
        pages = {25},
          doi = {10.1088/0004-637X/765/1/25},
archivePrefix = {arXiv},
       eprint = {1209.1392},
 primaryClass = {astro-ph.CO},
       adsurl = {https://ui.adsabs.harvard.edu/abs/2013ApJ...765...25N}
}

@ARTICLE{Montes2019,
       author = {{Montes}, Mireia and {Trujillo}, Ignacio},
        title = "{Intracluster light: a luminous tracer for dark matter in clusters of galaxies}",
      journal = {\mnras},
         year = 2019,
       volume = {482},
        pages = {2838-2851},
          doi = {10.1093/mnras/sty2858}
}

@ARTICLE{AlonsoAsensio2020,
       author = {{Alonso Asensio}, Isaac and {Dalla Vecchia}, Claudio and {Bah{\'e}}, Yannick M. and {Barnes}, David J. and {Kay}, Scott T.},
        title = "{The intracluster light as a tracer of the total matter density distribution: a view from simulations}",
      journal = {\mnras},
         year = 2020,
       volume = {494},
        pages = {1859-1864},
          doi = {10.1093/mnras/staa861}
}

@ARTICLE{Contini2021,
       author = {{Contini}, Emanuele},
        title = "{On the Origin and Evolution of the Intra-Cluster Light: A Brief Review of the Most Recent Developments}",
      journal = {Galaxies},
         year = 2021,
       volume = {9},
          eid = {60},
        pages = {60},
          doi = {10.3390/galaxies9030060}
}

@ARTICLE{Montes2022,
       author = {{Montes}, Mireia},
        title = "{The faint light in groups and clusters of galaxies}",
      journal = {Nature Astronomy},
         year = 2022,
       volume = {6},
        pages = {308-316},
          doi = {10.1038/s41550-022-01616-z}
}

@ARTICLE{Rudick2009,
       author = {{Rudick}, Craig S. and {Mihos}, J. Christopher and {Frey}, Lucille H. and {McBride}, Cameron K.},
        title = "{Tidal streams of intracluster light}",
      journal = {\apj},
         year = 2009,
       volume = {699},
        pages = {1518-1529},
          doi = {10.1088/0004-637x/699/2/1518}
}

@ARTICLE{Contini2014,
       author = {{Contini}, E. and {De Lucia}, G. and {Villalobos}, {\'A}. and {Borgani}, S.},
        title = "{On the formation and physical properties of the intracluster light in hierarchical galaxy formation models}",
      journal = {\mnras},
         year = 2014,
       volume = {437},
        pages = {3787-3802},
          doi = {10.1093/mnras/stt2174}
}

@ARTICLE{Willman2004,
       author = {{Willman}, Beth and {Governato}, Fabio and {Wadsley}, James and {Quinn}, Thomas},
        title = "{The origin and properties of intracluster stars in a rich cluster}",
      journal = {\mnras},
         year = 2004,
       volume = {355},
        pages = {159-168},
          doi = {10.1111/j.1365-2966.2004.08312.x}
}

@ARTICLE{Rudick2006,
       author = {{Rudick}, Craig S. and {Mihos}, J. Christopher and {McBride}, Cameron},
        title = "{The Formation and Evolution of Intracluster Light}",
      journal = {\apj},
         year = 2006,
       volume = {648},
        pages = {936-946},
          doi = {10.1086/506176}
}

@ARTICLE{Murante2007,
       author = {{Murante}, G. and {Giovalli}, M. and {Gerhard}, O. and {Arnaboldi}, M. and {Borgani}, S. and {Dolag}, K.},
        title = "{The importance of mergers for the origin of intracluster stars in cosmological simulations of galaxy clusters}",
      journal = {\mnras},
         year = 2007,
       volume = {377},
        pages = {2-16},
          doi = {10.1111/j.1365-2966.2007.11568.x}
}

@ARTICLE{Conroy2007,
       author = {{Conroy}, Charlie and {Wechsler}, Risa H. and {Kravtsov}, Andrey V.},
        title = "{The Hierarchical Build-Up of Massive Galaxies and the Intracluster Light since $z = 1$}",
      journal = {\apj},
         year = 2007,
       volume = {668},
        pages = {826-838},
          doi = {10.1086/521425}
}

@ARTICLE{Montes2014,
       author = {{Montes}, Mireia and {Trujillo}, Ignacio},
        title = "{Intracluster Light at the Frontier: A2744}",
      journal = {\apj},
         year = 2014,
       volume = {794},
          eid = {137},
        pages = {137},
          doi = {10.1088/0004-637x/794/2/137}
}

@ARTICLE{DeMaio2015,
       author = {{DeMaio}, Tahlia and {Gonzalez}, Anthony H. and {Zabludoff}, Ann and {Zaritsky}, Dennis and {Brada{\v c}}, Maru{\v s}a},
        title = "{On the origin of the intracluster light in massive galaxy clusters}",
      journal = {\mnras},
         year = 2015,
       volume = {448},
        pages = {1162-1177},
          doi = {10.1093/mnras/stv033}
}

@ARTICLE{Morishita2017,
       author = {{Morishita}, Takahiro and {Abramson}, Louis E. and {Treu}, Tommaso and {Schmidt}, Kasper B. and {Vulcani}, Benedetta and {Wang}, Xin},
        title = "{Characterizing Intracluster Light in the Hubble Frontier Fields}",
      journal = {\apj},
         year = 2017,
       volume = {846},
          eid = {139},
        pages = {139},
          doi = {10.3847/1538-4357/aa8403}
}

@ARTICLE{DeMaio2018,
       author = {{DeMaio}, Tahlia and {Gonzalez}, Anthony H. and {Zabludoff}, Ann and {Zaritsky}, Dennis and {Connor}, Thomas and {Donahue}, Megan and {Mulchaey}, John S.},
        title = "{Lost but not forgotten: intracluster light in galaxy groups and clusters}",
      journal = {\mnras},
         year = 2018,
       volume = {474},
        pages = {3009-3031},
          doi = {10.1093/mnras/stx2946}
}

@ARTICLE{Montes2018,
       author = {{Montes}, Mireia and {Trujillo}, Ignacio},
        title = "{Intracluster light at the Frontier - II. The Frontier Fields Clusters}",
      journal = {\mnras},
         year = 2018,
       volume = {474},
        pages = {917-932},
          doi = {10.1093/mnras/stx2847}
}

@ARTICLE{Feldmeier2004,
       author = {{Feldmeier}, John J. and {Mihos}, J. Christopher and {Morrison}, Heather L. and {Harding}, Paul and {Kaib}, Nathan and {Dubinski}, John},
        title = "{Deep CCD Surface Photometry of Galaxy Clusters. II. Searching for Intracluster Starlight in Non-cD clusters}",
      journal = {\apj},
         year = 2004,
       volume = {609},
        pages = {617-637},
          doi = {10.1086/421313}
}

@ARTICLE{Burke2015,
       author = {{Burke}, Claire and {Hilton}, Matt and {Collins}, Chris},
        title = "{Coevolution of brightest cluster galaxies and intracluster light using CLASH}",
      journal = {\mnras},
         year = 2015,
       volume = {449},
        pages = {2353-2367},
          doi = {10.1093/mnras/stv450}
}

@ARTICLE{Pillepich2018,
       author = {{Pillepich}, Annalisa and {Nelson}, Dylan and {Hernquist}, Lars and {Springel}, Volker and {Pakmor}, R{\"u}diger and {Torrey}, Paul and {Weinberger}, Rainer and {Genel}, Shy and {Naiman}, Jill P. and {Marinacci}, Federico and {Vogelsberger}, Mark},
        title = "{First results from the IllustrisTNG simulations: the stellar mass content of groups and clusters of galaxies}",
      journal = {\mnras},
         year = 2018,
       volume = {475},
        pages = {648-675},
          doi = {10.1093/mnras/stx3112}
}

@ARTICLE{Gonzalez2005,
       author = {{Gonzalez}, Anthony H. and {Zabludoff}, Ann I. and {Zaritsky}, Dennis},
        title = "{Intracluster Light in Nearby Galaxy Clusters: Relationship to the Halos of Brightest Cluster Galaxies}",
      journal = {\apj},
         year = 2005,
       volume = {618},
        pages = {195-213},
          doi = {10.1086/425896}
}

@ARTICLE{Zibetti2005,
       author = {{Zibetti}, S. and {White}, S.~D.~M. and {Schneider}, D.~P. and {Brinkmann}, J.},
        title = "{Intergalactic stars in $z \sim 0.25$ galaxy clusters: systematic properties from stacking of Sloan Digital Sky Survey imaging data}",
      journal = {\mnras},
         year = 2005,
       volume = {358},
        pages = {949-967},
          doi = {10.1111/j.1365-2966.2005.08817.x}
}

@ARTICLE{Kluge2020,
       author = {{Kluge}, M. and {Neureiter}, B. and {Riffeser}, A. and {Bender}, R. and {Goessl}, C. and {Hopp}, U. and {Schmidt}, M. and {Ries}, C. and {Brosch}, N.},
        title = "{Structure of Brightest Cluster Galaxies and Intracluster Light}",
      journal = {\apjs},
         year = 2020,
       volume = {247},
          eid = {43},
        pages = {43},
          doi = {10.3847/1538-4365/ab733b}
}

@ARTICLE{Joo2023,
       author = {{Joo}, Hyungjin and {Jee}, M. James},
        title = "{Intracluster light is already abundant at redshift beyond unity}",
      journal = {\nat},
         year = 2023,
       volume = {613},
        pages = {37-41},
          doi = {10.1038/s41586-022-05396-4}
}

@ARTICLE{Rudick2011,
       author = {{Rudick}, Craig S. and {Mihos}, J. Christopher and {McBride}, Cameron K.},
        title = "{The Quantity of Intracluster Light: Comparing Theoretical and Observational Measurement Techniques using Simulated Clusters}",
      journal = {\apj},
         year = 2011,
       volume = {732},
          eid = {48},
        pages = {48},
          doi = {10.1088/0004-637x/732/1/48}
}

@ARTICLE{Kluge2021,
       author = {{Kluge}, M. and {Bender}, R. and {Riffeser}, A. and {Goessl}, C. and {Hopp}, U. and {Schmidt}, M. and {Ries}, C.},
        title = "{Photometric Dissection of Intracluster Light and Its Correlations with Host Cluster Properties}",
      journal = {\apjs},
         year = 2021,
       volume = {252},
          eid = {27},
        pages = {27},
          doi = {10.3847/1538-4365/abcda6}
}

@ARTICLE{Contini2022,
       author = {{Contini}, E. and {Chen}, H.~Z. and {Gu}, Q.},
        title = "{The Transition Region between Brightest Cluster Galaxies and Intracluster Light in Galaxy Groups and Clusters}",
      journal = {\apj},
         year = 2022,
       volume = {928},
          eid = {99},
        pages = {99},
          doi = {10.3847/1538-4357/ac57c4}
}

@ARTICLE{Dolag2010,
       author = {{Dolag}, K. and {Murante}, G. and {Borgani}, S.},
        title = "{Dynamical difference between the cD galaxy and the diffuse, stellar component in simulated galaxy clusters}",
      journal = {\mnras},
         year = 2010,
       volume = {405},
        pages = {1544-1559},
          doi = {10.1111/j.1365-2966.2010.16583.x}
}

@ARTICLE{Cui2014,
       author = {{Cui}, Weiguang and {Murante}, G. and {Monaco}, P. and {Borgani}, S. and {Granato}, G.~L. and {Killedar}, M. and {De Lucia}, G. and {Presotto}, V. and {Dolag}, K.},
        title = "{Characterizing diffused stellar light in simulated galaxy clusters}",
      journal = {\mnras},
         year = 2014,
       volume = {437},
        pages = {816-830},
          doi = {10.1093/mnras/stt1940}
}

@ARTICLE{Remus2017,
       author = {{Remus}, Rhea-Silvia and {Dolag}, Klaus and {Hoffmann}, Tadziu},
        title = "{The Outer Halos of Very Massive Galaxies: BCGs and their DSC in the Magneticum Simulations}",
      journal = {Galaxies},
         year = 2017,
       volume = {5},
          eid = {49},
        pages = {49},
          doi = {10.3390/galaxies5030049}
}

@ARTICLE{Marini2024,
       author = {{Marini}, I. and {Saro}, A. and {Borgani}, S. and {Boi}, M.},
        title = "{Inferring intrahalo light from stellar kinematics}",
      journal = {\aap},
         year = 2024,
       volume = {689},
          eid = {A181},
        pages = {A181},
          doi = {10.1051/0004-6361/202449632}
}

@ARTICLE{Longobardi2015,
       author = {{Longobardi}, Alessia and {Arnaboldi}, Magda and {Gerhard}, Ortwin and {Hanuschik}, Reinhard},
        title = "{The outer regions of the giant Virgo galaxy M 87. Kinematic separation of stellar halo and intracluster light}",
      journal = {\aap},
         year = 2015,
       volume = {579},
          eid = {A135},
        pages = {A135},
          doi = {10.1051/0004-6361/201525773}
}

@ARTICLE{Hartke2018,
       author = {{Hartke}, J. and {Arnaboldi}, M. and {Gerhard}, O. and {Agnello}, A. and {Longobardi}, A. and {Coccato}, L. and {Pulsoni}, C. and {Freeman}, K.~C. and {Merrifield}, M.},
        title = "{Three dynamically distinct stellar populations in the halo of M49}",
      journal = {\aap},
         year = 2018,
       volume = {616},
          eid = {A123},
        pages = {A123},
          doi = {10.1051/0004-6361/201832711}
}

@ARTICLE{Dressler1979,
       author = {{Dressler}, A.},
        title = "{The dynamics and structure of the cD galaxy in Abell 2029}",
      journal = {\apj},
         year = 1979,
       volume = {231},
          eid = {659},
        pages = {659},
          doi = {10.1086/157229}
}

@ARTICLE{Kelson2002,
       author = {{Kelson}, Daniel D. and {Zabludoff}, Ann I. and {Williams}, K.~A. and {Trager}, S.~C. and {Mulchaey}, J.~S. and {Bolte}, Michael},
        title = "{Determination of the Dark Matter Profile of A2199 from Integrated Starlight}",
      journal = {\apj},
         year = 2002,
       volume = {576},
        pages = {720-737},
          doi = {10.1086/341891}
}

@ARTICLE{Loubser2008,
       author = {{Loubser}, S.~I. and {Sansom}, A.~E. and {S{\'a}nchez-Bl{\'a}zquez}, P. and {Soechting}, I.~K. and {Bromage}, G.~E.},
        title = "{Radial kinematics of brightest cluster galaxies}",
      journal = {\mnras},
         year = 2008,
       volume = {391},
        pages = {1009-1028},
          doi = {10.1111/j.1365-2966.2008.13813.x}
}

@ARTICLE{Loubser2018,
       author = {{Loubser}, S.~I. and {Hoekstra}, H. and {Babul}, A. and {O'Sullivan}, E.},
        title = "{Diversity in the stellar velocity dispersion profiles of a large sample of brightest cluster galaxies $z \leq 0.3$}",
      journal = {\mnras},
         year = 2018,
       volume = {477},
        pages = {335-358},
          doi = {10.1093/mnras/sty498}
}

@ARTICLE{Edwards2020,
       author = {{Edwards}, Louise O. V. and {Salinas}, Matthew and {Stanley}, Steffanie and {Holguin West}, Priscilla E. and {Trierweiler}, Isabella and {Alpert}, Hannah and {Coelho}, Paula and {Koppaka}, Saisneha and {Tremblay}, Grant R. and {Martel}, Hugo and {Li}, Yuan},
        title = "{Clocking the formation of today's largest galaxies: wide field integral spectroscopy of brightest cluster galaxies and their surroundings}",
      journal = {\mnras},
         year = 2020,
       volume = {491},
        pages = {2617-2638},
          doi = {10.1093/mnras/stz2706}
}

@ARTICLE{Bender2015,
       author = {{Bender}, Ralf and {Kormendy}, John and {Cornell}, Mark E. and {Fisher}, David B.},
        title = "{Structure and Formation of cD Galaxies: NGC 6166 in Abell 2199}",
      journal = {\apj},
         year = 2015,
       volume = {807},
          eid = {56},
        pages = {56},
          doi = {10.1088/0004-637x/807/1/56}
}

@ARTICLE{Barbosa2018,
       author = {{Barbosa}, C.~E. and {Arnaboldi}, M. and {Coccato}, L. and {Gerhard}, O. and {Mendes de Oliveira}, C. and {Hilker}, M. and {Richtler}, T.},
        title = "{Sloshing in its cD halo: MUSE kinematics of the central galaxy NGC 3311 in the Hydra I cluster}",
      journal = {\aap},
         year = 2018,
       volume = {609},
          eid = {A78},
        pages = {A78},
          doi = {10.1051/0004-6361/201731834}
}

@ARTICLE{Hilker2018,
       author = {{Hilker}, M. and {Richtler}, T. and {Barbosa}, C.~E. and {Arnaboldi}, M. and {Coccato}, L. and {Mendes de Oliveira}, C.},
        title = "{The Hydra I cluster core. II. Kinematic complexity in a rising velocity dispersion profile around the cD galaxy NGC 3311}",
      journal = {\aap},
         year = 2018,
       volume = {619},
          eid = {A70},
        pages = {A70},
          doi = {10.1051/0004-6361/201731737}
}

@ARTICLE{Planck2020,
       author = {{Planck Collaboration}},
        title = "{Planck 2018 results. VI. Cosmological parameters}",
      journal = {\aap},
         year = 2020,
       volume = {641},
          eid = {A6},
        pages = {A6},
          doi = {10.1051/0004-6361/201833910}
}

@ARTICLE{Jee2010,
       author = {{Jee}, M.~J.},
        title = "{Tracing the Peculiar Dark Matter Structure in the Galaxy Cluster Cl 0024+17 with Intracluster Stars and Gas}",
      journal = {\apj},
         year = 2010,
       volume = {717},
        pages = {420-434},
          doi = {10.1088/0004-637X/717/1/420}
}

@ARTICLE{Joo2026,
       author = {{Joo}, Hyungjin and {Jee}, M. James and {Finner}, Kyle and {Scofield}, Zachary P. and {Cha}, Sangjun and {Kim}, Jinhyub and {Chary}, Ranga-Ram and {Faisst}, Andreas and {Lee}, Bomee},
        title = "{Mature but Still Growing: JWST Detection of the Earliest Intracluster Light at $z \sim 2$}",
      journal = {\apjl},
         year = 2026,
       volume = {1002},
          eid = {L17},
        pages = {L17},
          doi = {10.3847/2041-8213/ae5c9f}
}

@ARTICLE{Joo2025,
       author = {{Joo}, Hyungjin and {Jee}, M. James and {Kim}, Juhan and {Lee}, Jaehyun and {Ko}, Jongwan and {Park}, Changbom and {Shin}, Jihye and {Snaith}, Owain and {Pichon}, Christophe and {Gibson}, Brad and {Kim}, Yonghwi},
        title = "{Tracing the Formation History of Intrahalo Light with Horizon Run 5}",
      journal = {\apj},
         year = 2025,
       volume = {990},
          eid = {96},
        pages = {96},
          doi = {10.3847/1538-4357/adf4d0}
}

@ARTICLE{Yoo2024,
       author = {{Yoo}, Jaewon and {Park}, Changbom and {Sabiu}, Cristiano G. and {Singh}, Ankit and {Ko}, Jongwan and {Lee}, Jaehyun and {Pichon}, Christophe and {Jee}, M. James and {Gibson}, Brad K. and {Snaith}, Owain and {Kim}, Juhan and {Shin}, Jihye and {Kim}, Yonghwi and {Kim}, Hyowon},
        title = "{Spatial Distribution of Intracluster Light versus Dark Matter in Horizon Run 5}",
      journal = {\apj},
         year = 2024,
       volume = {965},
          eid = {145},
        pages = {145},
          doi = {10.3847/1538-4357/ad2df8}
}

@ARTICLE{Mercurio2021,
       author = {{Mercurio}, A. and {Rosati}, P. and {Biviano}, A. and {Annunziatella}, M. and {Girardi}, M. and {Sartoris}, B. and {Nonino}, M. and {Brescia}, M. and {Riccio}, G. and {Grillo}, C. and {Balestra}, I. and {Caminha}, G. B. and {De Lucia}, G. and {Gobat}, R. and {Seitz}, S. and {Tozzi}, P. and {Scodeggio}, M. and {Vanzella}, E. and {Angora}, G. and {Bergamini}, P. and {Borgani}, S. and {Demarco}, R. and {Meneghetti}, M. and {Strazzullo}, V. and {Tortorelli}, L. and {Umetsu}, K. and {Fritz}, A. and {Gruen}, D. and {Kelson}, D. and {Lombardi}, M. and {Maier}, C. and {Postman}, M. and {Rodighiero}, G. and {Ziegler}, B.},
        title = "{CLASH-VLT: Abell S1063. Cluster assembly history and spectroscopic catalogue}",
      journal = {\aap},
         year = 2021,
       volume = {656},
          eid = {A147},
        pages = {A147},
          doi = {10.1051/0004-6361/202142168}
}

@ARTICLE{Biviano2013,
       author = {{Biviano}, A. and {Rosati}, P. and {Balestra}, I. and {Mercurio}, A. and {Girardi}, M. and {Nonino}, M. and {Grillo}, C. and {Scodeggio}, M. and {Lemze}, D. and {Kelson}, D. and {Umetsu}, K. and {Postman}, M. and {Zitrin}, A. and {Czoske}, O. and {Ettori}, S. and {Fritz}, A. and {Lombardi}, M. and {Maier}, C. and {Medezinski}, E. and {Mei}, S. and {Presotto}, V. and {Strazzullo}, V. and {Tozzi}, P. and {Ziegler}, B. and {Annunziatella}, M. and {Bartelmann}, M. and {Benitez}, N. and {Bradley}, L. and {Brescia}, M. and {Broadhurst}, T. and {Coe}, D. and {Demarco}, R. and {Donahue}, M. and {Ford}, H. and {Gobat}, R. and {Graves}, G. and {Koekemoer}, A. and {Kuchner}, U. and {Melchior}, P. and {Meneghetti}, M. and {Merten}, J. and {Moustakas}, L. and {Munari}, E. and {Reg{\H o}s}, E. and {Sartoris}, B. and {Seitz}, S. and {Zheng}, W.},
        title = "{CLASH-VLT: The mass, velocity-anisotropy, and pseudo-phase-space density profiles of the $z=0.44$ galaxy cluster MACS J1206.2-0847}",
      journal = {\aap},
         year = 2013,
       volume = {558},
          eid = {A1},
        pages = {A1},
          doi = {10.1051/0004-6361/201321955}
}

@ARTICLE{Girardi2015,
       author = {{Girardi}, M. and {Mercurio}, A. and {Balestra}, I. and {Nonino}, M. and {Biviano}, A. and {Grillo}, C. and {Rosati}, P. and {Annunziatella}, M. and {Demarco}, R. and {Fritz}, A. and {Gobat}, R. and {Lemze}, D. and {Presotto}, V. and {Scodeggio}, M. and {Tozzi}, P. and {Bartosch Caminha}, G. and {Brescia}, M. and {Coe}, D. and {Kelson}, D. and {Koekemoer}, A. and {Lombardi}, M. and {Medezinski}, E. and {Postman}, M. and {Sartoris}, B. and {Umetsu}, K. and {Zitrin}, A. and {Boschin}, W. and {Czoske}, O. and {De Lucia}, G. and {Kuchner}, U. and {Maier}, C. and {Meneghetti}, M. and {Monaco}, P. and {Monna}, A. and {Munari}, E. and {Seitz}, S. and {Verdugo}, M. and {Ziegler}, B.},
        title = "{CLASH-VLT: Substructure in the galaxy cluster MACS J1206.2-0847 from kinematics of galaxy populations}",
      journal = {\aap},
         year = 2015,
       volume = {579},
          eid = {A4},
        pages = {A4},
          doi = {10.1051/0004-6361/201425599}
}

@ARTICLE{Presotto2014,
       author = {{Presotto}, V. and {Girardi}, M. and {Nonino}, M. and {Mercurio}, A. and {Grillo}, C. and {Rosati}, P. and {Biviano}, A. and {Annunziatella}, M. and {Balestra}, I. and {Cui}, W. and {Sartoris}, B. and {Lemze}, D. and {Ascaso}, B. and {Moustakas}, J. and {Ford}, H. and {Fritz}, A. and {Czoske}, O. and {Ettori}, S. and {Kuchner}, U. and {Lombardi}, M. and {Maier}, C. and {Medezinski}, E. and {Molino}, A. and {Scodeggio}, M. and {Strazzullo}, V. and {Tozzi}, P. and {Ziegler}, B. and {Bartelmann}, M. and {Benitez}, N. and {Bradley}, L. and {Brescia}, M. and {Broadhurst}, T. and {Coe}, D. and {Donahue}, M. and {Gobat}, R. and {Graves}, G. and {Kelson}, D. and {Koekemoer}, A. and {Melchior}, P. and {Meneghetti}, M. and {Merten}, J. and {Moustakas}, L. A. and {Munari}, E. and {Postman}, M. and {Reg{\H o}s}, E. and {Seitz}, S. and {Umetsu}, K. and {Zheng}, W. and {Zitrin}, A.},
        title = "{Intracluster light properties in the CLASH-VLT cluster MACS J1206.2-0847}",
      journal = {\aap},
         year = 2014,
       volume = {565},
          eid = {A126},
        pages = {A126},
          doi = {10.1051/0004-6361/201323251}
}

@ARTICLE{Postman2012,
       author = {{Postman}, Marc and {Coe}, Dan and {Ben{\'\i}tez}, Narciso and {Bradley}, Larry and {Broadhurst}, Tom and {Donahue}, Megan and {Ford}, Holland and {Graur}, Or and {Graves}, Genevieve and {Jouvel}, Stephanie and {Koekemoer}, Anton and {Lemze}, Doron and {Medezinski}, Elinor and {Molino}, Alberto and {Moustakas}, Leonidas and {Ogaz}, Sara and {Riess}, Adam and {Rodney}, Steve and {Rosati}, Piero and {Umetsu}, Keiichi and {Zheng}, Wei and {Zitrin}, Adi and {Bartelmann}, Matthias and {Bouwens}, Rychard and {Czakon}, Nicole and {Golwala}, Sunil and {Host}, Ole and {Infante}, Leopoldo and {Jha}, Saurabh and {Jimenez-Teja}, Yolanda and {Kelson}, Daniel and {Lahav}, Ofer and {Lazkoz}, Ruth and {Maoz}, Dani and {McCully}, Curtis and {Melchior}, Peter and {Meneghetti}, Massimo and {Merten}, Julian and {Moustakas}, John and {Nonino}, Mario and {Patel}, Brandon and {Reg{\"o}s}, Enik{\"o} and {Sayers}, Jack and {Seitz}, Stella and {Van der Wel}, Arjen},
        title = "{The Cluster Lensing and Supernova Survey with Hubble: An Overview}",
      journal = {\apjs},
         year = 2012,
       volume = {199},
          eid = {25},
        pages = {25},
          doi = {10.1088/0067-0049/199/2/25}
}

@ARTICLE{Lotz2017,
       author = {{Lotz}, J. M. and {Koekemoer}, A. and {Coe}, D. and {Grogin}, N. and {Capak}, P. and {Mack}, J. and {Anderson}, J. and {Avila}, R. and {Barker}, E. A. and {Borncamp}, D. and {Brammer}, G. and {Durbin}, M. and {Gunning}, H. and {Hilbert}, B. and {Jenkner}, H. and {Khandrika}, H. and {Levay}, Z. and {Lucas}, R. A. and {MacKenty}, J. and {Ogaz}, S. and {Porterfield}, B. and {Reid}, N. and {Robberto}, M. and {Royle}, P. and {Smith}, L. J. and {Storrie-Lombardi}, L. J. and {Sunnquist}, B. and {Surace}, J. and {Taylor}, D. C. and {Williams}, R. and {Bullock}, J. and {Dickinson}, M. and {Finkelstein}, S. and {Natarajan}, P. and {Richard}, J. and {Robertson}, B. and {Tumlinson}, J. and {Zitrin}, A. and {Flanagan}, K. and {Sembach}, K. and {Soifer}, B. T. and {Mountain}, M.},
        title = "{The Frontier Fields: Survey Design and Initial Results}",
      journal = {\apj},
         year = 2017,
       volume = {837},
          eid = {97},
        pages = {97},
          doi = {10.3847/1538-4357/837/1/97}
}

@ARTICLE{Weilbacher2020,
       author = {{Weilbacher}, Peter M. and {Palsa}, Ralf and {Streicher}, Ole and {Bacon}, Roland and {Urrutia}, Tanya and {Wisotzki}, Lutz and {Conseil}, Simon and {Husemann}, Bernd and {Jarno}, Aur{\'e}lien and {Kelz}, Andreas and {P{\'e}contal-Rousset}, Arlette and {Richard}, Johan and {Roth}, Martin M. and {Selman}, Fernando and {Vernet}, Jo{\"e}l},
        title = "{The data processing pipeline for the MUSE instrument}",
      journal = {\aap},
         year = 2020,
       volume = {641},
          eid = {A28},
        pages = {A28},
          doi = {10.1051/0004-6361/202037855}
}

@ARTICLE{Bertin1996,
       author = {{Bertin}, E. and {Arnouts}, S.},
        title = "{SExtractor: Software for source extraction}",
      journal = {\aaps},
         year = 1996,
       volume = {117},
        pages = {393-404},
          doi = {10.1051/aas:1996164}
}

@ARTICLE{Cappellari2004,
       author = {{Cappellari}, Michele and {Emsellem}, Eric},
        title = "{Parametric Recovery of Line-of-Sight Velocity Distributions from Absorption-Line Spectra of Galaxies via Penalized Likelihood}",
      journal = {\pasp},
         year = 2004,
       volume = {116},
        pages = {138-147},
          doi = {10.1086/381875}
}

@ARTICLE{Cappellari2017,
       author = {{Cappellari}, Michele},
        title = "{Improving the full spectrum fitting method: accurate convolution with Gauss-Hermite functions}",
      journal = {\mnras},
         year = 2017,
       volume = {466},
        pages = {798-811},
          doi = {10.1093/mnras/stw3020}
}

@ARTICLE{Cappellari2023,
       author = {{Cappellari}, Michele},
        title = "{Full spectrum fitting with photometry in pPXF: stellar population versus dynamical masses, non-parametric star formation history and metallicity for 3200 LEGA-C galaxies at redshift $z\approx0.8$}",
      journal = {\mnras},
         year = 2023,
       volume = {526},
        pages = {3273-3300},
          doi = {10.1093/mnras/stad2597}
}

@ARTICLE{Vazdekis2016,
       author = {{Vazdekis}, A. and {Koleva}, M. and {Ricciardelli}, E. and {R{\"o}ck}, B. and {Falc{\'o}n-Barroso}, J.},
        title = "{UV-extended E-MILES stellar population models: young components in massive early-type galaxies}",
      journal = {\mnras},
         year = 2016,
       volume = {463},
        pages = {3409-3436},
          doi = {10.1093/mnras/stw2231}
}

@ARTICLE{Pietrinferni2004,
       author = {{Pietrinferni}, Adriano and {Cassisi}, Santi and {Salaris}, Maurizio and {Castelli}, Fiorella},
        title = "{A Large Stellar Evolution Database for Population Synthesis Studies. I. Scaled Solar Models and Isochrones}",
      journal = {\apj},
         year = 2004,
       volume = {612},
        pages = {168-190},
          doi = {10.1086/422498}
}

@ARTICLE{Pietrinferni2006,
       author = {{Pietrinferni}, Adriano and {Cassisi}, Santi and {Salaris}, Maurizio and {Castelli}, Fiorella},
        title = "{A Large Stellar Evolution Database for Population Synthesis Studies. II. Stellar Models and Isochrones with an $\alpha$-enhanced Metal Distribution}",
      journal = {\apj},
         year = 2006,
       volume = {642},
        pages = {797-812},
          doi = {10.1086/501344}
}

@ARTICLE{Feroz2008,
       author = {{Feroz}, F. and {Hobson}, M. P.},
        title = "{Multimodal nested sampling: an efficient and robust alternative to Markov Chain Monte Carlo methods for astronomical data analyses}",
      journal = {\mnras},
         year = 2008,
       volume = {384},
        pages = {449-463},
          doi = {10.1111/j.1365-2966.2007.12353.x}
}

@ARTICLE{Feroz2009,
       author = {{Feroz}, F. and {Hobson}, M. P. and {Bridges}, M.},
        title = "{MultiNest: an efficient and robust Bayesian inference tool for cosmology and particle physics}",
      journal = {\mnras},
         year = 2009,
       volume = {398},
        pages = {1601-1614},
          doi = {10.1111/j.1365-2966.2009.14548.x}
}

@ARTICLE{Feroz2019,
       author = {{Feroz}, Farhan and {Hobson}, Michael P. and {Cameron}, Ewan and {Pettitt}, Anthony N.},
        title = "{Importance Nested Sampling and the MultiNest Algorithm}",
      journal = {The Open Journal of Astrophysics},
         year = 2019,
       volume = {2},
          eid = {10},
        pages = {10},
          doi = {10.21105/astro.1306.2144}
}

@ARTICLE{Cappellari2003,
       author = {{Cappellari}, M. and {Copin}, Y.},
        title = "{Adaptive spatial binning of integral-field spectroscopic data using Voronoi tessellations}",
      journal = {\mnras},
         year = 2003,
       volume = {342},
        pages = {345-354},
          doi = {10.1046/j.1365-8711.2003.06541.x}
}

@ARTICLE{Buchner2014,
       author = {{Buchner}, J. and {Georgakakis}, A. and {Nandra}, K. and {Hsu}, L. and {Rangel}, C. and {Brightman}, M. and {Merloni}, A. and {Salvato}, M. and {Donley}, J. and {Kocevski}, D.},
        title = "{X-ray spectral modelling of the AGN obscuring region in the CDFS: Bayesian model selection and catalogue}",
      journal = {\aap},
         year = 2014,
       volume = {564},
          eid = {A125},
        pages = {A125},
          doi = {10.1051/0004-6361/201322971}
}

@ARTICLE{Fusco2020,
       author = {{Fusco}, T. and {Bacon}, R. and {Kamann}, S. and {Conseil}, S. and {Neichel}, B. and {Correia}, C. and {Beltramo-Martin}, O. and {Vernet}, J. and {Kolb}, J. and {Madec}, P.-Y.},
        title = "{Reconstruction of the ground-layer adaptive-optics point spread function for MUSE wide field mode observations}",
      journal = {\aap},
         year = 2020,
       volume = {635},
          eid = {A208},
        pages = {A208},
          doi = {10.1051/0004-6361/202037595}
}

@ARTICLE{Speagle2020,
       author = {{Speagle}, Joshua S},
        title = "{dynesty: a dynamic nested sampling package for estimating Bayesian posteriors and evidences}",
      journal = {\mnras},
         year = 2020,
       volume = {493},
        pages = {3132-3158},
          doi = {10.1093/mnras/staa278}
}

@ARTICLE{Navarro1997,
       author = {{Navarro}, Julio F. and {Frenk}, Carlos S. and {White}, Simon D. M.},
        title = "{A Universal Density Profile from Hierarchical Clustering}",
      journal = {\apj},
         year = 1997,
       volume = {490},
        pages = {493-508},
          doi = {10.1086/304888}
}

@ARTICLE{Bryan1998,
       author = {{Bryan}, Greg L. and {Norman}, Michael L.},
        title = "{Statistical Properties of X-Ray Clusters: Analytic and Numerical Comparisons}",
      journal = {\apj},
         year = 1998,
       volume = {495},
        pages = {80-99},
          doi = {10.1086/305262}
}

@ARTICLE{Sersic1963,
       author = {{S{\'e}rsic}, J. L.},
        title = "{Influence of the atmospheric and instrumental dispersion on the brightness distribution in a galaxy}",
      journal = {Boletin de la Asociacion Argentina de Astronomia La Plata Argentina},
         year = 1963,
       volume = {6},
        pages = {41-43},
       adsurl = {https://ui.adsabs.harvard.edu/abs/1963BAAA....6...41S}
}

@BOOK{BinneyTremaine2008,
       author = {{Binney}, James and {Tremaine}, Scott},
        title = "{Galactic Dynamics: Second Edition}",
    publisher = {Princeton University Press},
      address = {Princeton, NJ},
         year = 2008
}

@MISC{Bradley2024,
       author = {{Bradley}, Larry and {Sip{\H o}cz}, Brigitta and {Robitaille}, Thomas and {Tollerud}, Erik and {Vin{\'\i}cius}, Z{\`e} and {Deil}, Christoph and {Barbary}, Kyle and {Wilson}, Tom J. and {Busko}, Ivo and {Donath}, Axel and {G{\"u}nther}, Hans Moritz and {Cara}, Mihai and {Lim}, Pey Lian and {Me{\ss}linger}, Stefan and {Burnett}, Zach and {Conseil}, Simon and {Droettboom}, Michael and {Bostroem}, Azalee and {Bray}, E. M. and {Bratholm}, Lars Andersen and {Jamieson}, William and {Ginsburg}, Adam and {Barentsen}, Geert and {Craig}, Matt and {Rathi}, Shivani and {Pascual}, Sergio and {Perren}, Gabriel and {Georgiev}, Iskren Y. and {de Val-Borro}, Miguel and {Kerzendorf}, Wolfgang and {Bach}, Yoonsoo P. and {Quint}, Brett and {Souchereau}, Harrison},
        title = "{astropy/photutils}",
         year = 2024,
    publisher = {Zenodo},
          doi = {10.5281/zenodo.596036}
}

@ARTICLE{Zwicky1951,
       author = {{Zwicky}, F.},
        title = "{The Coma Cluster of Galaxies}",
      journal = {\pasp},
         year = 1951,
        month = apr,
       volume = {63},
        pages = {61},
          doi = {10.1086/126318}
}

@ARTICLE{Ko2018,
       author = {{Ko}, Jongwan and {Jee}, M. James},
        title = "{Evidence for the Existence of Abundant Intracluster Light at z = 1.24}",
      journal = {\apj},
         year = 2018,
        month = aug,
       volume = {862},
       number = {2},
          eid = {95},
        pages = {95},
          doi = {10.3847/1538-4357/aacbda},
archivePrefix = {arXiv},
       eprint = {1806.02687},
 primaryClass = {astro-ph.GA}
}

@ARTICLE{Cha2025,
       author = {{Cha}, Sangjun and {Cho}, Boseong Young and {Joo}, Hyungjin and {Lee}, Wonki and {HyeongHan}, Kim and {Scofield}, Zachary P. and {Finner}, Kyle and {Jee}, M. James},
        title = "{A High-caliber View of the Bullet Cluster through JWST Strong and Weak Lensing Analyses}",
      journal = {\apjl},
         year = 2025,
       volume = {987},
       number = {1},
          eid = {L15},
        pages = {L15},
          doi = {10.3847/2041-8213/add2f0},
archivePrefix = {arXiv},
       eprint = {2503.21870},
 primaryClass = {astro-ph.CO}
}

@ARTICLE{Lee2026,
       author = {{Lee}, Wonki and {Nelson}, Dylan and {Jee}, M. James and {Despali}, Giulia and {Vogelsberger}, Mark and {Pillepich}, Annalisa and {Roche}, Cian and {Finner}, Kyle and {Joo}, Hyungjin and {Nagai}, Daisuke and {ZuHone}, John},
        title = "{TNGC-SIDM I: Evolution of massive cluster mergers under the self-interacting dark matter paradigm}",
      journal = {arXiv e-prints},
         year = 2026,
        month = oct,
          eid = {arXiv:2610.18957},
        pages = {arXiv:2610.18957},
archivePrefix = {arXiv},
       eprint = {2610.18957}
}

@ARTICLE{Jee2026,
       author = {{Jee}, M. James and {Park}, Dongak and {Lee}, Wonki},
        title = "{A New Robust Constraint on the Self-interaction Cross Section of Dark Matter with Double Radio Relic Clusters}",
      journal = {\apj},
         year = 2026,
        month = aug,
       volume = {1007},
       number = {1},
          eid = {56},
        pages = {56},
          doi = {10.3847/1538-4357/ae88fb},
archivePrefix = {arXiv},
       eprint = {2605.00093}
}
\bibliographystyle{aasjournalv7}

\end{document}